\def\anonflag{false}\def\reviewflag{false}\def\subtitleflag{false}\def\epigraphflag{true}\def\acksflag{false}
\PassOptionsToPackage{dvipsnames}{xcolor}
\providecommand{\anonflag}{true}
\providecommand{\reviewflag}{true}
\providecommand{\subtitleflag}{true}
\providecommand{\epigraphflag}{false}
\providecommand{\acksflag}{true}
\def\paperflagtrue{true}
\edef\paperclassoptions{acmsmall,screen%
  \ifx\reviewflag\paperflagtrue,review\fi
  \ifx\anonflag\paperflagtrue,anonymous\fi}
\expandafter\documentclass\expandafter[\paperclassoptions]{acmart}

\usepackage{hyperref}       %
\usepackage{url}            %
\usepackage{amsfonts}       %
\usepackage[dvipsnames]{xcolor}         %
\usepackage{colortbl}       %
\usepackage{subcaption}
\usepackage{framed}         %
\usepackage{multirow}
\usepackage{latexsym}

\usepackage{xspace}
\usepackage{thmtools}
\usepackage{thm-restate}
\usepackage{cleveref}
\usepackage{pifont}        %

\DeclareMathSizes{10}{8}{6}{5}
\Crefformat{figure}{#2Fig.~#1#3}
\Crefformat{section}{\S#2#1#3}
\Crefformat{subsection}{\S#2#1#3}
\Crefformat{subsubsection}{\S#2#1#3}
\usepackage{wrapfig}

\newcommand\codecomment[1]{\textcolor{codecomment-color}{\#~#1}}

\usepackage{pgfplots}
\pgfplotsset{compat=1.18}
\usepgfplotslibrary{groupplots}
\usepackage{pgfplotstable}   %

\definecolor{keyword-color}{HTML}{000000}
\definecolor{keyword-color-1}{HTML}{9465B6}
\definecolor{keyword-color-2}{HTML}{9465B6}
\definecolor{codecomment-color}{HTML}{10ac84}
\definecolor{string-color}{HTML}{AA99CC}
\usepackage{listings, lstautogobble}
\usepackage[customcolors,shade]{hf-tikz}
\usepackage{tikz}
  \usetikzlibrary{babel,arrows.meta,backgrounds,positioning,shapes.symbols,decorations.pathreplacing,decorations.pathmorphing,shadows.blur,tikzmark}
\usepackage{tikz-cd}    %

\usepackage{stackengine}

\theoremstyle{plain}

\newtheorem{theorem}{Theorem}[section]

\newtheorem{lemma}[theorem]{Lemma}

\usepackage[inline]{enumitem}

\setlist[enumerate]{leftmargin=5.5mm}
\setlist[itemize]{leftmargin=4mm}
\setlist[description]{font=\normalfont}   %

\usepackage{mathtools}

\usepackage{mathpartir}     %
\makeatletter
{\small\xdef\mpr@smallsize{\f@size}}
\def\mpr@labelsize{\ifdim\f@size pt>\mpr@smallsize pt \small\fi}
\let\atmostsmall\mpr@labelsize

\def\lefttir@name#1{\hbox{\mpr@labelsize\LeftTirNameStyle{#1}}}
\def\righttir@name#1{\hbox{\mpr@labelsize\RightTirNameStyle{#1}}}
\let\TirName\lefttir@name
\let\LeftTirName\lefttir@name
\let\DefTirName\lefttir@name
\let\LabTirName\lefttir@name
\let\RightTirName\righttir@name
\makeatother
\usepackage{stmaryrd}

\usepackage{ottalt}

  \newcommand{\ottdrule}[4][]{{\displaystyle\frac{\begin{array}{l}#2\end{array}}{#3}\quad\ottdrulename{#4}}}

\newcommand{\ottpremise}[1]{ #1 \\}

\newcommand{\ottdrulename}[1]{\textsc{#1}}

\newcommand{\ottafterlastrule}{\\}

\newcommand{\ottdruleTXXDo}[1]{\ottdrule[#1]{%
    \ottpremise{(op: in \to out)}%
  }{ \Env \vdash \dokw{}\, op: in \to out}{%
 {\ottdrulename{T-Do}}{}%
 }}

\newcommand{\ottdruleTXXResume}[1]{\ottdrule[#1]{%
    \ottpremise{\resumekw{}: \sigma \to \tau \in \Env}%
  }{ \Env \vdash \resumekw{}: \sigma \to \tau}{%
 {\ottdrulename{T-Resume}}{}%
 }}

\newcommand{\ottdruleTXXHandler}[1]{\ottdrule[#1]{%
    \ottpremise{\Env \vdash \overline{a} : par}%
    \ottpremise{\Env \vdash e : \tau}%
    \ottpremise{\Env \vdash j: par,\tau \to \tau'}%
    \ottpremise{\Env,\kw{resume}:par, out_i \to \tau'  \vdash j_i: par,in_i\to  \tau'}%
    \ottpremise{(op_i : in_i \to out_i, i = 1 \ldots n)}
  }{ \Env \vdash \defhandleret{\overline{a}}{e}{j}{op_1 \mapsto j_1,\dots, op_n
      \mapsto j_n } : \unit \to \tau'}{%
 {\ottdrulename{T-Handler}}{}%
 }}

  \renewottcommands[ott]

\renewcommand\ottaltinferrule[4]{
  \inferrule*[narrower=1,lab=#1,#2]
    {#3}
    {#4}
}
\renewcommand\MathparLineskip{\lineskip=0cm}
\makeatletter
{\footnotesize\xdef\mpr@fnsize{\f@size}}
\def\atleastfootnotesize{\ifdim\f@size pt<\mpr@fnsize pt \footnotesize\fi}
\makeatother
\renewcommand*\FormatDruleSectionHead[1]{{\atleastfootnotesize\fbox{#1}}}
\renewcommand*\FormatDruleSectionHeads[1]{{\atleastfootnotesize\fbox{\strut#1}}}
\renewcommand*\FormatDruleSectionHeadRight[1]{{\atleastfootnotesize\emph{(#1)}}}

\newcommand{\newrule}[1]{{\atmostsmall\textsc{#1}}}

\makeatletter
\newcommand{\rulelabel}[2]{%
  \edef\@currentlabel{#1}%
  \phantomsection
  \label{#2}%
}
\makeatother

\makeatletter
\newcommand{\customlabel}[2]{%
   \phantomsection%
   \edef\@currentlabel{#1}%
   \protected@write\@auxout{}{\string\newlabel{#2}{{#1}{\thepage}{#1}{\@currentHref}{}}}%
}
\makeatother

\renewcommand\rrefstyle{\atmostsmall\normalfont\scshape}
\newcommand{\ruleref}[1]{{\atmostsmall\textsc{\ref{#1}}}}

\graphicspath{ {./} }

\setcopyright{none}
\renewcommand\footnotetextcopyrightpermission[1]{}
\usepackage[para]{footmisc} %
 
\begin{document}

\clearpage{}%

\newcommand{\Rcut}[1]{#1}
\newcommand{\myif}[3]
 {\mathbf{if}\, #1\, \mathbf{then} \, #2 \, \mathbf{else}\, #3}
 \newcommand{\NF}{\mathrm{NF}}
\newcommand{\qqquad}{\qquad \quad}

\def\code{\texttt}
\newcommand{\cons}[2]{#1 \!::\! #2}
\newcommand{\true}{\mathrm{t}}
\newcommand{\false}{\mathrm{f}}
\newcommand{\jaxfx}{\texttt{effecax}\xspace}
\newcommand{\jaxfxppl}{\ensuremath{\jaxfx_{\texttt{PPL}}}\xspace}
\newcommand{\machinespecs}{an NVIDIA A100 (SXM4, 40\,GB) GPU\xspace}
\newcommand{\compiletimespec}{a 12-vCPU Intel Xeon (Cascade Lake, 2.20\,GHz) host with 85\,GB RAM\xspace}
\newcommand{\minijaxpr}{\ensuremath{\lambda_{\textsc{Jaxpr}}}\xspace}
\newcommand{\minijaxprfx}{\ensuremath{\lambda^{\epsilon}_{\textsc{Jaxpr}}}\xspace}

\newcommand{\handletransform}{\codepy{handle_effects}}  %
\newcommand{\testfootnotesize}{\atmostsmall}
\newcommand{\kw}[1]{\ensuremath{\mathkw{{#1}}}}

\newcommand{\kww}[1]{\textcolor{gray}{\testfootnotesize\ensuremath{\mathkw{#1}}}}

\newcommand{\lambdakw}{\kw{lambda}}
\newcommand{\lambdasym}{$\lambda$}
\newcommand{\integer}{\mathbb{Z}}
\newcommand{\float}{\mathbb{R}}
\newcommand{\bool}{\mathbb{B}}

\let\oldcomma\,

\newcommand{\defcond}[2]{\ensuremath{\mathkw{cond}\,[\,{\kww{then:}}\,#1,\,\kww{else:}\,#2\,]}}
\newcommand{\defgeneralcond}[1]{\ensuremath{\mathkw{cond}\,[\,{\kww{branches:}}\,#1]}}
\newcommand{\defwhile}[2]{\ensuremath{\mathkw{while}\,[\,\kww{cond:}\,#1,\,\kww{body:}\,#2\,]}}
\newcommand{\defscan}[2]{\ensuremath{\mathkw{scan}\,[\,\kww{body:}\,#1,\,\kww{carry:}\,#2]}}
\newcommand{\defcall}[1]{\ensuremath{\mathkw{call}\,[\,\kww{body:}\,#1\,]}}
\newcommand{\defdo}[1]{\ensuremath{\mathkw{do}\;#1}}
\newcommand{\defhandleret}[4]{\ensuremath{\mathkw{handle}\,[\,\kww{args:}\,#1,\,\kww{body:}\,#2,\,\kww{ret:}\,#3,\,\kww{impl:}\,#4]}}
\newcommand{\defhandle}[3]{\ensuremath{\mathkw{handle}\,[\,\kww{args:}\,#1,\,\kww{body:}\,#2,\,\kww{impl:}\,#3]}}
\newcommand{\defletnobody}[3]{\ensuremath{\mathkw{let}\;#1\;\mathkw{=}\;#2\;#3}}
\newcommand{\deflet}[4]{\ensuremath{\mathkw{let}\;#1\;\mathkw{=}\;#2\;#3\;\mathkw{in}\;#4}}
\newcommand{\defreturn}[1]{\ensuremath{\mathkw{return}\;#1}}
\newcommand{\deflambda}[2]{\ensuremath{\{\,\lambda#1.\,#2\,\}}}

\newcommand\bind[1]{\mathbin{\vcenter{\hbox{$\scriptstyle\rhd$}}} {#1}.}

\definecolor{hcolor}{HTML}{3F6E8F}
\definecolor{parcolor}{HTML}{5F6B7A}
\definecolor{kcolor}{HTML}{9A4F2E}
\definecolor{tcolor}{HTML}{000000}   %
\definecolor{trcolor}{HTML}{000000}
\newcommand\ccont[1]{{\color{kcolor}#1}}
\newcommand\cpar[1]{{\color{parcolor}#1}}
\newcommand\chandler[1]{{\color{hcolor}#1}}
\newcommand\ctype[1]{{\color{gray}#1}}

\newcommand{\judgsize}{\footnotesize}
\makeatletter
\newcommand{\judg}[1]{\text{{\ifdim\f@size pt>\mpr@smallsize pt \judgsize\fi $#1$}}}
\makeatother

\newcommand{\jsep}{\mathrel{\raisebox{0.3ex}{\scalebox{0.8}[0.55]{$\vert$}}}}
\newcommand{\newtransformeh}[4]
{\judg{\mathcolor{trcolor}{\mathcal{E}}\lBrack \, {\color{hcolor} #1} \jsep {\color{parcolor} #3} \jsep  {\color{kcolor} #2} \vdash  #4 \, \rBrack}}
\newcommand{\newtransformjh}[4]
{\judg{\mathcolor{trcolor}{\mathcal{J}}\lBrack \, {\color{hcolor} #1} \jsep {\color{parcolor} #3} \jsep  {\color{kcolor} #2} \vdash  #4 \, \rBrack}}

\newcommand{\newtransformehip}[4]
{\judg{\mathcolor{trcolor}{\mathcal{E}_{\mathit{ip}}}\lBrack \, {\color{hcolor} #1} \jsep {\color{parcolor} #3} \jsep  {\color{kcolor} #2} \vdash  #4 \, \rBrack}}

\newcommand{\newtransformehnopar}[3]
{\judg{\mathcal{E}\lBrack \, {\color{hcolor} #1} \jsep {\color{kcolor} #2}  \vdash  #3 \, \rBrack}}

\newcommand{\transformeh}[2]{\mathcal{T}\!e_{\,#1}\lBrack{#2}\rBrack}
\newcommand{\transformjh}[2]{\mathcal{T}_{\,#1}\lBrack{#2}\rBrack}

\newcommand{\delay}[2]{\mathcal{D}\!e_{\,#1}\lBrack{#2}\rBrack}

\newcommand{\newdelay}[4]
{\judg{\mathcal{D}\lBrack \, {\color{hcolor} #1} \jsep {\color{parcolor} #3} \jsep {\color{kcolor} #2} \vdash  #4 \, \rBrack}}

\newcommand*{\mathkw}[1]{\ensuremath{\texttt{\textbf{#1}}}}
\newcommand*{\mathmono}[1]{\ensuremath{\texttt{#1}}}
\newcommand*{\mathbrace}[1]{\mathmono{\{}~#1~\mathmono{\}}}

\newcommand*{\seq}[1]{\ensuremath{\overline{#1}}}

\newcommand{\gordoncut}[1]{}

\newcommand{\letkw}{\kw{let}}
\newcommand{\letreckw}{\kw{letrec}}
\newcommand{\dokw}{\kw{do}}
\newcommand{\resumekw}{\kw{resume}}
\newcommand{\inkw}{\kw{in}}
\newcommand{\eqkw}{\kw{=}}
\newcommand{\retkw}{\kw{return}}
\newcommand{\mreturn}[2]{\kw{return}_{#1}\; #2}
\newcommand{\handlekw}{\kw{handle}}
\newcommand{\condkw}{\kw{cond}}
\newcommand{\whilekw}{\kw{while}}

\newcommand{\nattype}{\mathbb{N}}
\newcommand{\dtype}{d}
\newcommand{\shape}{s}
\newcommand{\booltype}{\mathkw{bool}}
\newcommand{\floattype}{\mathkw{real}}
\newcommand{\inttype}{\mathkw{int}}
\newcommand{\unittype}{\mathkw{unit}}
\newcommand\unit{\star}

\newcommand{\replus}{\mathkw{+}}
\newcommand{\restar}{{}^{\mathkw{*}}}
\newcommand{\remaybe}{\mathkw{?}}

\newcommand*{\ctxhole}{\langle \cdot \rangle}
\newcommand*{\rvctx}[1]{E\langle {#1} \rangle}
\newcommand*{\lvctx}[1]{E_{\mathit{lv}}\langle {#1} \rangle}

\newcommand{\Env}{\Gamma}
\newcommand{\evalto}{\longrightarrow}
\newcommand{\rwcond}{\mathrel{\leadsto_{1}}}
\newcommand{\rwabort}{\mathrel{\leadsto_{2}}}
\newcommand{\rwcong}{\mathrel{\leadsto^{*}_{2}}}
\newcommand{\condsim}{\mathrel{\simeq_{\mathit{cond}}}}
\newcommand{\abortsim}{\mathrel{\simeq_{\mathit{abort}}}}
\newcommand{\refines}{\lesssim}

\newcommand{\ponders}[3]{\marginpar{\tiny\itshape\raggedright\textcolor{#2}{\textbf{#1:} #3}}\ignorespaces} \marginparwidth=1.4cm \marginparsep=0cm

\newcommand{\dan}[1]{\ponders{Dan}{blue}{#1}}
\newcommand{\shangyin}[1]{\ponders{Shangyin}{orange}{#1}}
\newcommand{\ningning}[1]{{\ponders{Ningning}{magenta}{#1}}}
\newcommand{\gordon}[1]{\ponders{Gordon}{cyan}{#1}}
\newcommand{\gaga}[1]{\ponders{Gaga}{purple}{#1}}

\lstset{mathescape}  %

\makeatletter
\newcommand{\tinyy}{\@setfontsize{\tinyy}{5pt}{5pt}}
\makeatother

\newcommand{\reducedstrut}{\vrule width 0pt height .9\ht\strutbox depth .9\dp\strutbox\relax}
\newcommand{\highlightblue}[1]{%
  \begingroup
  \setlength{\fboxsep}{0pt}%
  \colorbox{blue!20}{\reducedstrut#1\/}%
  \endgroup
}
\newcommand{\highlightgreen}[1]{%
  \begingroup
  \setlength{\fboxsep}{0pt}%
  \colorbox{green!20}{\reducedstrut#1\/}%
  \endgroup
}
\newcommand{\highlightorange}[1]{%
  \begingroup
  \setlength{\fboxsep}{0pt}%
  \colorbox{orange!40}{\reducedstrut#1\/}%
  \endgroup
}
\newcommand{\highlightyellow}[1]{%
  \begingroup
  \setlength{\fboxsep}{0pt}%
  \colorbox{yellow!40}{\reducedstrut#1\/}%
  \endgroup
}
\newcommand{\highlightred}[1]{%
  \begingroup
  \setlength{\fboxsep}{0pt}%
  \colorbox{red!20}{\reducedstrut#1\/}%
  \endgroup
}
\newcommand{\highlightgray}[1]{%
 \begingroup
 \setlength{\fboxsep}{0pt}%
 \colorbox{gray!30}{\reducedstrut#1\/}%
 \endgroup
}

\newcommand{\codejaxpr}[1]{{#1}}

\lstnewenvironment{ttlisting}{\lstset{basicstyle=\ttfamily}}{}

\definecolor{codegreen}{rgb}{0,0.6,0}
\definecolor{codegray}{rgb}{0.5,0.5,0.5}
\definecolor{codepurple}{rgb}{0.58,0,0.82}
\definecolor{backcolour}{rgb}{0.95,0.95,0.92}
\definecolor{wingreen}{HTML}{D4EFDF} %

\lstdefinestyle{python}{
    language=Python,
    inputencoding=utf8,
    extendedchars=true,  %
    commentstyle=\color{codegreen},
    keywordstyle=\color{magenta},
    numberstyle=\tiny\color{codegray},
    stringstyle=\color{codepurple},
    basicstyle=\ttfamily\tiny,
    breakatwhitespace=false,
    breaklines=true,
    captionpos=b,
    keepspaces=true,
    morekeywords={with, as},
    literate={%
      {set}{{set}}{3}
      {θ}{{\texttheta}}{1}
      {→}{{\rightarrow}}{1}
    },
    showspaces=false,
    showstringspaces=false,
    showtabs=false,
    tabsize=2,
    xleftmargin=1em, %
    breaklines=true,
    postbreak=\mbox{\textcolor{red}{$\hookrightarrow$}\space},
    escapechar=!,
}

\lstdefinestyle{jaxpr}{
    inputencoding=utf8,
    extendedchars=true,  %
    commentstyle=\color{codegreen},
    keywordstyle=\bfseries\color{magenta},
    numberstyle=\tiny\color{codegray},
    stringstyle=\color{codepurple},
    basicstyle=\ttfamily\tiny,
    breakatwhitespace=false,
    breaklines=true,
    captionpos=b,
    keepspaces=true,
    alsoletter={=>[]},   
    morekeywords={handle, call_k, =>, return, call, in, lambda, let},
    showspaces=false,
    showstringspaces=false,
    showtabs=false,
    tabsize=2,
    breaklines=true,
    postbreak=\mbox{\textcolor{red}{$\hookrightarrow$}\space},
    escapechar=!,
}

\lstdefinestyle{effects}{
    inputencoding=utf8,
    extendedchars=true,  %
    commentstyle=\color{codegreen},
    keywordstyle={\sffamily\bfseries},
    numberstyle=\tiny\color{codegray},
    stringstyle=\color{codepurple},
    basicstyle=\ttfamily\footnotesize,
    breakatwhitespace=false,
    breaklines=true,
    captionpos=b,
    keepspaces=true,
    alsoletter={=>[]},   
    morekeywords={handle,return,perform},
    otherkeywords={:}, %
    literate={%
      {->}{{$\to$}}{2}
      {<-}{{$\leftarrow$}}{2}
      {|->}{{$\mapsto$}}{2}
      {lambda}{{$\lambda$}}{1}
    },
    showspaces=false,
    showstringspaces=false,
    showtabs=false,
    tabsize=2,
    breaklines=true,
    postbreak=\mbox{\textcolor{red}{$\hookrightarrow$}\space},
    escapechar=!,
}

\newcommand{\codefx}[1]{\lstinline[style=effects,basicstyle=\ttfamily\small]{#1}}

\newcommand{\gcut}[1]{}

\newcommand{\sms}{\overline{\sigma}}
\newcommand{\lift}{\mathrm{lift}}
\newcommand{\myeqdef}{=_{\small \mbox{def}}} %
\newcommand{\transformd}[2]{\mathcal{T}_{#1}[\!| #2 |\!]}
\renewcommand{\mreturn}{\mathrm{mreturn}}
\newcommand{\mbind}{\mathrm{mbind}}

\newcommand{\jaxapp}[3]{#1 {\scriptscriptstyle \color{gray}@}_{#2} #3}
\newcommand{\defapp}[2]{#1\, #2}

\newcommand\nulle[1]{ \langle #1 \rangle }

\newcommand{\Is}{\judg{{\color{hcolor}\mathbb{H}}}}
\newcommand{\hs}{\mathbb{H}}
\newcommand{\ts}{\overline{\tau}}
\newcommand{\ks}{\judg{{\color{kcolor}\mathbb{K}}}}
\newcommand{\ps}{\judg{{\color{parcolor}\mathbb{P}}}}
\newcommand{\as}{\overline{a}}
\newcommand{\cs}{\overline{c}}
\newcommand{\bs}{\overline{b}}
\newcommand{\ls}{\overline{l}}
\newcommand{\js}{\overline{j}}
\newcommand{\ws}{\overline{w}}
\newcommand{\xs}{\overline{x}}
\newcommand{\ys}{\overline{y}}
\newcommand{\zs}{\overline{z}}
\newcommand{\us}{\overline{u}}
\newcommand{\lps}{\overline{p}}
\newcommand{\pars}{\overline{par}}

\newcommand\spec[1]{ {\color{Orchid}\langle}  \ccont{#1} {\color{Orchid}\rangle} }

\newcommand{\que}{\mathord{?}}
\newcommand{\eC}{[\;]}

\newcommand{\newTR}[4]
{\judg{{\color{tcolor}\mathcal{R}}\lBrack \, {\color{hcolor} #1} \jsep {\color{parcolor} #2} \jsep {\color{gray} #3} \vdash  #4 \, \rBrack}}
\newcommand{\TR}[2]{\mathcal{R}_{#1}[\![#2]\!]}

\newcommand{\gordoncutextended}[1]{}

\newcommand\safediv{\code{safe\_div}\xspace}
\newcommand\throw{\code{throw}}
\newcommand\getop{\code{get}}
\newcommand\setop{\code{set}}

\newcommand{\prem}[1]{{\footnotesize$#1$}}

\makeatletter
\newcommand{\dispsize}[1]{\edef\dispsize@bls{\the\baselineskip}#1\baselineskip=\dispsize@bls\relax}
\makeatother

\newcommand{\badmark}{(\textcolor{red!70!black}{\ding{55}})}
\newcommand{\okmark}{(\textcolor{green!45!black}{\ding{51}})}

\providecommand\myeqSim[1]{\,\;\simeq^{#1}\,}
\providecommand\simcps[1]{{\color{kcolor}#1}}
\providecommand\siminl[1]{{\color{hcolor}#1}}
\newcommand\tail[1]{\rfloor \, #1 \, \lfloor}
\newlength{\tailrulew}
\newcommand{\trule}[2]{\makebox[\tailrulew][l]{\newrule{(#1)}}\customlabel{#1}{#2}}
\newcommand{\econt}{\hspace{20pt}\phantom{=\;\enskip}}
\newcommand{\leanlogo}{\includegraphics[height=\fontcharht\font`\L]{figures/logos/lean-logo}}
\newcommand{\leanproof}{\hyperref[tab:lean]{\leanlogo}}
\makeatletter
\newcommand{\thmnoteorlean}[1]{\@thmnoteorlean#1\@nil{#1}}
\def\@thmnoteorlean#1#2\@nil#3{%
  \ifx#1\leanproof
    \@ifnotempty{#2}{ {\the\thm@notefont(#2)}}%
  \else
    \ {\the\thm@notefont(#3)}%
  \fi}
\def\thmhead@plain#1#2#3{%
  \thmname{#1}\thmnumber{\@ifnotempty{#1}{ }\@upn{#2}}%
  \thmnote{\thmnoteorlean{#3}}}
\let\thmhead\thmhead@plain
\makeatother
\clearpage{}%

\renewcommand\Coloneqq{\mathrel{\mathop{:\!:}}=}

\title{Handling JAX}
\ifx\subtitleflag\paperflagtrue
\subtitle{Compiling Effect Handlers to a Restricted, First-Order Target}
\fi

\author{Oghenevwogaga Ebresafe}
\orcid{0009-0002-1951-7424}
\email{gaga@cs.toronto.edu}
\affiliation{%
  \institution{University of Toronto}
  \country{Canada}
}

\author{Daniel Zheng}
\orcid{0009-0006-4523-6262}
\email{danielzheng@google.com}
\affiliation{%
  \institution{Google DeepMind}
  \country{USA}
}

\author{Gordon Plotkin}
\orcid{0000-0001-8496-6096}
\email{plotkin@google.com}
\affiliation{%
  \institution{Google DeepMind}
  \country{USA}
}

\author{Ningning Xie}
\orcid{0000-0002-5961-1493}
\email{ningningxie@google.com}
\affiliation{%
  \institution{Google DeepMind and University of Toronto}
  \country{Canada}
}

\begin{abstract}
  Languages that target hardware accelerators typically expose a restricted,
  first-order core without recursion or first-class functions, leaving
  programmers without a principled way to express computational effects. This
  paper presents a technique for compiling effect handlers to such a constrained
  target. We introduce a novel translation that statically eliminates effect
  handlers, and prove that it preserves both typing and semantics. Moreover, we
  address the challenges of integrating effect handlers with structured control
  flow, where continuations cannot be constructed statically. The calculus and
  all our results are fully mechanized in the Lean 4 proof assistant. Finally,
  we implement our design as a library for JAX, a Python domain-specific
  language for accelerator programming, and demonstrate its expressiveness
  through practical case studies.
\end{abstract}

\maketitle

\section{Introduction}

\ifx\epigraphflag\paperflagtrue
\begingroup
\noindent\textit{``When walking about the countryside of Italy, the people will not hesitate to tell you that JAX has `una anima di pura programmazione funzionale'.''}\par
\noindent\quad -- JAX documentation\footnote{\url{https://docs.jax.dev/en/latest/notebooks/Common_Gotchas_in_JAX.html}}\par
\endgroup
\vspace{6pt}
\fi
Languages that target hardware accelerators are deliberately restrictive. To
generate efficient code for GPUs and TPUs, they typically expose a purely
functional, \emph{first-order} core: programs are built from array primitives
and a fixed set of structured control-flow constructs such as conditionals and
loops, without recursion or first-class functions. Examples include array
languages such as Futhark~\cite{henriksen2017futhark},
Dex~\cite{paszke2021getting}, and Accelerate~\cite{chakravarty2011accelerating},
and compiler intermediate representations such as XLA~\cite{xla}. While these
restrictions enable aggressive optimization, they leave programmers without a
principled way to express computational effects such as mutable state and
exceptions. Without language support, each such feature is typically implemented
via a bespoke program transformation or an external library.

We study this problem in the setting of
{JAX}%
~\cite{jax2018github}, a
\textit{domain-specific language} embedded in Python for accelerator-oriented
array computations, widely adopted in high-performance numerical computing and
large-scale machine learning, including for Google's \textit{large language
  models}~\cite{palm,cloudtpu2023training,team2023gemini} and by
Anthropic~\cite{Anthropic2026Transparency}. JAX traces Python programs into a
first-order intermediate representation, \emph{jaxprs}, with exactly the
restrictions described above, and lacks direct support for effectful features.
As a result, popular JAX libraries resort to different, often \textit{ad-hoc}
mechanisms to implement these features, causing inconsistencies and
interoperability issues across the ecosystem (see \Cref{sec:overview}).

We address this challenge using \textit{algebraic effect
  handlers}~\cite{algeff2003,effecthandlers2009}, a flexible and composable
mechanism for defining structured control-flow abstractions. While effect
handlers have been incorporated into various
languages~\cite{bauer2015programming, koka2017,ocamleffects2021}, including
recent work on low-level languages such as C~\cite{ceffects2024,leijen2017c},
C++~\cite{cppeffects2022}, and WebAssembly~\cite{webassembly2023}, all such
implementations rely on features that first-order languages lack: higher-order
functions, recursion, or a capturable runtime stack. The central question of
this paper is therefore: \emph{how can effect handlers be compiled to a
  first-order target without higher-order functions or recursion?} We resolve
this by introducing a translation that statically eliminates handlers, carefully
extending the technique across loop boundaries where continuations cannot be
constructed statically.

\paragraph{{Contributions}}

Following the high-level overview in \Cref{sec:overview},
we present our contributions in the following sections:
\begin{itemize}
\item \textbf{Formalization} (\Cref{sec:calculus}): We formalize
  \minijaxprfx, a calculus that extends a core subset of JAX's intermediate
  representation, a first-order language with second-order control-flow
  primitives, with \textit{parameterized algebraic effect handlers}.
  We establish preservation and progress with effects.
\item \textbf{Translation} (\Cref{sec:effect-translation}):
  We introduce a novel continuation-based translation that statically compiles
  effect handlers away without needing higher-order functions or recursion,
  and prove that it preserves both typing and semantics.
\item \textbf{Iterative control flow} (\Cref{subsec:tail-resumptive-effects}):
  We analyze the challenges of translating effect handlers across loop
  boundaries, where continuations cannot be constructed statically, and present
  a mechanism to support \textit{tail-resumptive} and \textit{abortive} handlers while
  preserving handler parameters and scoping.
\item \textbf{Mechanization}: We fully mechanize our calculus, its type system
  and operational semantics, translation, and all soundness proofs in the Lean 4
  proof assistant.
\item \textbf{Implementation} (\Cref{sec:implementation}): We implement
  \jaxfx{}, an effect handler library for JAX, outlining its programming
  interface and our implementation.
\item \textbf{Evaluation} (\Cref{sec:case-studies}): We evaluate \jaxfx{} across
  practical case studies, demonstrating that it provides an expressive,
  flexible, and natural interface for JAX programs.
\end{itemize}

\noindent
While this paper focuses on JAX, where \jaxfx provides immediate utility to a
broad user base, neither the problem nor our solution is specific to it: the
compilation applies to any first-order target, such as XLA's intermediate
representation. More broadly, we hope this work offers a new perspective on
compiling effect handlers to low-level target languages. Lastly, we discuss
related work in \Cref{sec:related}.

\section{Overview}
\label{sec:overview}

\subsection{Motivation}
\label{subsec:motivation}

JAX adopts a functional programming paradigm: its compilation pipeline is
designed for pure Python functions free of side effects. This model aligns
naturally with modern machine learning workloads, which predominantly consist of
static mathematical operations connected in computation graphs. Functional
purity simplifies graph optimization and yields significant performance
gains~\cite{frostig2019compiling}. Moreover, purity enables JAX to
reliably \emph{trace} Python functions into its intermediate representation,
\emph{jaxpr}\footnote{\url{https://docs.jax.dev/en/latest/jaxpr.html}}: a
first-order, statically typed language in A-normal form~\cite{FlanaganANF}, with
a fixed set of second-order control-flow primitives such as \code{cond} and
\code{while}. JAX also provides powerful function transformations,
such as
\code{jax.grad} for automatic differentiation,
\texttt{jax.vmap} for automatic vectorization,
and \texttt{jax.jit} for just-in-time compilation
of functions to hardware accelerators via XLA~\cite{xla}.

The effectiveness of this functional programming model is evidenced by JAX's
widespread adoption in the machine learning
community~\cite{palm,cloudtpu2023training,team2023gemini,Anthropic2026Transparency}.
Its paradigm of composable transformations is so powerful that it inspired
PyTorch’s
\texttt{functorch}\footnote{\url{https://github.com/pytorch/functorch}}, which
has been fully integrated into core
PyTorch\footnote{\url{https://docs.pytorch.org/docs/2.14/func.html}}.

However, while the functional programming model is highly desirable for machine
learning workloads, practical ML programming is inherently stateful and
side-effect intensive. ML training pipelines must manage mutable state, handle
exceptions, perform asynchronous computation, and occasionally use specialized
effects such as coroutines in probabilistic programming. In short,
computational effects abound.
This creates a fundamental tension: JAX’s functional core provides no principled
mechanism for needed side-effectful computations. To illustrate this, we
examine two effects common in everyday ML programming: mutable state and
exception handling.

\paragraph{\textbf{Case 1: Mutable state}}

To use mutable references, developers can either leverage JAX’s experimental
built-in \texttt{Ref}
arrays\footnote{\url{https://docs.jax.dev/en/latest/101/state.html}} or rely
on external libraries. The \texttt{Ref} approach uses a custom program
transformation that converts programs containing \texttt{Ref}s into purely
functional programs by automatically threading state parameters through the
computation.

Alternatively, JAX-based libraries such as \texttt{Flax
  NNX}\footnote{\url{https://github.com/google/flax}}, a popular neural network
library, expose \code{Module}s that are mutable by default. To maintain
compatibility with JAX’s functional requirements, \texttt{NNX} implements its
own transformations, such as \code{flax.nnx.jit}. This transformation wraps
\code{jax.jit} by ``lifting’’ the mutable module: it splits the module into a
static structure and a pure state, applies the transformation to those pure
components, and then propagates any state updates back to the original mutable
object.

Both approaches involve trade-offs. While \texttt{Ref} integrates tightly with
JAX’s built-in transformations, it is constrained by JAX’s tracing rules: for
instance, \texttt{Ref}s cannot be returned from \code{jit}-compiled functions
or from the bodies of higher-order primitives (e.g.~\code{while}), and cannot
be passed where arrays are expected. Conversely, \code{flax.nnx.jit} provides a
more ergonomic mutable interface, but requires a duplicated ecosystem of
custom transformations and incurs Python performance overhead in training loops from
repeatedly lifting stateful
objects\footnote{\url{https://flax.readthedocs.io/en/stable/guides/performance.html}}.

\paragraph{\textbf{Case 2: Exceptions}}

A similar situation exists for exception handling. JAX provides an experimental
built-in transformation called
\texttt{checkify}\footnote{\url{https://docs.jax.dev/en/latest/debugging/checkify\_guide.html}},
which rewrites a program that may raise errors into a functionally pure version
that returns error values. JAX libraries such as
\texttt{Equinox}\footnote{\url{https://github.com/patrick-kidger/equinox}}, on
the other hand, offer \texttt{equinox.error\_if} to raise runtime errors
imperatively during execution.

While \texttt{checkify} adheres to JAX’s functional paradigm and composes
cleanly with other transformations, it introduces significant boilerplate by
forcing developers to manually propagate and inspect error values throughout
the computation. It also lacks support for early termination, %
which can be
important for performance.
Conversely, \texttt{Equinox} makes it easy to insert
error checks into existing models without refactoring, but degrades performance
by forcing synchronous execution barriers between the hardware accelerator and
the host.

\paragraph{\textbf{Current landscape and limitations}}

As a result, JAX programmers must constantly choose between specialized, ad-hoc
JAX transformations or fragmented external libraries. Neither is ideal.
Dedicated internal transformations encourage code duplication and lack
modularity. For instance, the \texttt{Ref} transformation implements state
mutation using a substantial amount of code, while the \texttt{checkify}
transformation relies on an entirely separate implementation. Despite this, both
transformations share similar structures and ``parameter plumbing'' patterns. This
redundancy inflates JAX's codebase and its surface for compiler bugs.

Alternatively, relying on custom libraries often results in performance
degradation, poor tool integration, and ecosystem fragmentation.
Interoperability remains a painful challenge. For example, disparate state
management mechanisms across libraries require complex, fragile manual state
conversions\footnote{As an example, see state conversions between \texttt{Haiku} and
  \texttt{Flax}\\
  \url{https://dm-haiku.readthedocs.io/en/latest/notebooks/flax.html\#Initialization-is-different}}.
Moreover, combining multiple effects from different libraries may not work~as~expected.

Furthermore, if a developer requires a specific effect for which no custom JAX
transformation or third-party library exists, they are forced to sacrifice
expressive power or resort to brittle, ad-hoc workarounds. Given JAX’s central
role in machine learning, the lack of a unified approach to computational
effects remains a fundamental
limitation.

\subsection{Key Ideas}
\label{sec:key-ideas}

To resolve these limitations, we seek a side-effect framework that has:
\begin{itemize}
\item \textit{Extensibility}: Developers should be able to define custom side
  effects as well as specify their corresponding handling logic. Multiple
  effects should compose seamlessly; for example, a program should be able to
  manage mutable state while simultaneously throwing exceptions.
\item \textit{Integration}: The mechanism should integrate seamlessly with
  the language's existing facilities. Moreover, since inspecting the
  intermediate representation is standard practice for debugging and
  optimization, it should produce clean, idiomatic code in that
  representation.
\item \textit{Backend preservation}: Accelerator-oriented languages rely on a
  downstream compiler, XLA in the case of JAX, to target hardware. The
  mechanism should remain fully compatible with existing backends and require
  no changes to that compiler.
\end{itemize}

In this work, we propose to extend JAX with \textit{algebraic effect
  handlers}~\cite{algeff2001, plotkin2013handling}, a flexible framework for
extensible and composable computational effects in functional settings.
Concretely, we extend the jaxpr representation with effect-handler constructs.
As an example, consider the function \safediv{}, which performs a $\throw$
operation on division by zero (types are omitted; the syntax is detailed in
\Cref{sec:calculus}):
\[\mbox{\footnotesize\setlength{\arraycolsep}{1pt}$
\begin{array}{@{}l@{\;}l@{\;}l@{}}
  \safediv \triangleq \{\, \lambda m, n. &
    \multicolumn{2}{@{}l@{}}{\deflet{b}{\code{eq}}{n~0}{}} \\
  & \mathkw{let}\; r = \mathkw{cond}\,[ &
      \kww{then:}\, \deflambda{x, y}{\deflet{z}{\dokw~\throw}{\unit}{\retkw~z}}, \\
  & & \kww{else:}\, \deflambda{x, y}{\deflet{z}{\code{div}}{x~y}{\retkw~z}}
      \,]~b~m~n \\
  & \multicolumn{2}{@{}l@{}}{\mathkw{in}\; \retkw~r \,\}}
\end{array}$}\]%
We handle it with a handler whose parameter ($\kww{args}$) is unused and thus
takes the unit value $\unit$, runs the
handled computation ($\kww{body}$), which calls \safediv{} on $42$ and $0$,
returns its result unchanged ($\kww{ret}$), and handles $\throw$ by
resuming the computation with the fallback value $0$ ($\kww{impl}$):
\[\mbox{\footnotesize\setlength{\arraycolsep}{1pt}$
\begin{array}{@{}l@{}}
    \mathkw{handle}\,[\, \kww{args:}\,\unit,\;
      \kww{body:}\, \deflet{w}{\defcall{\safediv}}{42~0}{\retkw~w},\;
      \kww{ret:}\, \deflambda{x, y}{\retkw~y}, \\
  \hspace{3.5em}
      \kww{impl:}\, \throw \mapsto
        \deflambda{x, y}{\deflet{u}{\resumekw}{\unit~0}{\retkw~u}} \,]~\unit
\end{array}$}\]%
We then compile these constructs away by a translation that produces
standard, vanilla jaxprs, ensuring that existing transformations and backends apply
unchanged. Realizing this plan, however, raises two key challenges, which we
discuss next.

\paragraph{\textbf{Challenge 1: a first-order target}}

Effect handler implementations typically capture operation continuations either
at runtime, by copying or manipulating the
stack~\cite{ocamleffects2021,ceffects2024,cppeffects2022,webassembly2023}, or at
compile time, via continuation passing style (CPS) translation that represents
continuations as
closures~\cite{hillerstrom2017continuation,iteratedcps2020,cpseffecthandlers2022}.
Jaxprs offer neither facility, as they have no runtime stack to capture, and no
higher-order functions or recursion to represent continuations.

We present a novel compilation strategy for this setting; to our knowledge, this
is the first work to study effect handlers under such restrictions.
Conceptually, our compilation combines a CPS transformation, which compiles
handlers into continuations, with \emph{partial evaluation} to statically
evaluate all handler-related constructs. As a result, neither handlers nor
continuations appear in the output, leaving only primitive computations for the
hardware. Fusing CPS with static reduction is a classic idea underlying one-pass
CPS transformations~\cite{danvy1992representing,danvy2003first} and CPS
translations of effect handlers without administrative
redexes~\cite{hillerstrom2017continuation,iteratedcps2020}; we apply it here
directly to the handler constructs themselves. On the example above, the
translation produces the following jaxpr:
\[\mbox{\footnotesize\setlength{\arraycolsep}{1pt}$
\begin{array}{@{}l@{\;}l@{\;}l@{}}
  \mathkw{let}\; z_1 = \mathkw{call}\,[\, \kww{body:}\, \{\, \lambda m, n. &
    \multicolumn{2}{@{}l@{}}{\deflet{b}{\code{eq}}{n~0}{}} \\
  & \mathkw{let}\; r = \mathkw{cond}\,[ &
      \kww{then:}\, \deflambda{x, y}{\deflet{u}{\defcall{\deflambda{p, z}{\retkw~z}}}{\unit~0}{\retkw~u}}, \\
  & & \kww{else:}\, \deflambda{x, y}{\deflet{z}{\code{div}}{x~y}{\retkw~z}}
      \,]~b~m~n \\
  & \multicolumn{2}{@{}l@{}}{\mathkw{in}\; \retkw~r \,\} \,]~42~0 \;\;
      \mathkw{in}\; \retkw~z_1}
\end{array}$}\]%
Here, the $\dokw~\throw$ in the $\kww{then}$ branch has been statically
evaluated to its handling code, here just the value $0$, while the primitive
computations are left untouched for execution on the hardware.

\paragraph{\textbf{Challenge 2: loop boundaries}}
This strategy, however, requires the continuation of every operation to be
known statically. Consider an operation performed inside a $\kw{while}$ loop
but handled \emph{outside} it, as in \Cref{fig:challenge-while}.
The continuation of the operation comprises not only the rest of the current
iteration but also all remaining iterations, that is, the loop itself. Translating
it would
then unroll
\begin{wrapfigure}{r}{0.44\textwidth}
\hspace{8pt}
{\footnotesize
$\arraycolsep=1pt
 \begin{array}{lllll}
   \multicolumn{5}{l}{\!\!\!\!\!\!\letkw{}~x:\sigma = \kw{handle} \, [}\\
                                        \kww{args:} & \multicolumn{3}{l}{\unit, \, \kww{ret:} \, j_r, \, \kww{impl:} \, \tikzmarknode{opcase}{op \mapsto j},  } \\
                                        \kww{body:} & \letkw{}~b = \kw{while}\, [ & \kww{cond:} & j_{1},  \\
                                                    & & \kww{body:} & \kw{let}~z = \tikzmarknode{opuse}{\dokw{}~op~a}\\
                                                    & &             & \kw{in}
                                                                      \ldots
                                                                      \, ] ~\as \\
                                                    & \inkw{}~ \ldots \, ] ~\unit\\
   \multicolumn{5}{l}{\!\!\!\!\!\!\inkw{}~\retkw~x}\\
 \end{array}
$
}%
\begin{tikzpicture}[overlay, remember picture]
  \draw[-{Stealth[length=1.6mm]}, semithick, gray!60!black]
    ([xshift=2pt]opuse.east) -- ++(0.3,0) |- ([xshift=4pt]opcase.east);
\end{tikzpicture}%
\vspace{3pt}
   \caption{Handling across the loop boundary}
  \label{fig:challenge-while}
  \vspace{2pt}
\end{wrapfigure}%
the loop. As a result, the translation does not terminate.
Unfortunately, this is a common use case:
the state and exception effects of \Cref{subsec:motivation} are typically performed
inside training and solver loops whose handlers live outside. We support these
most common cases by identifying two classes of handlers that can
nevertheless be compiled across loop boundaries.

We focus on \emph{tail-resumptive} handlers, which resume the computation
exactly once at the end of the operation clause and are common in
practice~\cite{xie2020effect,xie2021generalized}. For such common clauses, we
avoid materializing continuations entirely by handling operations
\emph{in-place} and inlining the clause directly into the loop body. This
approach introduces two subtleties that we systematically address: threading
handler parameters through the loop via an extension we call \emph{the final
  continuation}, and preventing naive inlining from altering the handler scope.
Compared to prior work, our approach is novel in several aspects: it applies to
jaxprs, supports parameterized handlers, and preserves semantics without
restrictions~\cite{xie2020effect,xie2021generalized}.

Additionally, we support \emph{abortive} handlers, which never resume and are
common for exceptions. We achieve this through a complementary source-level
rewrite of the $\kw{while}$ expression, denoted $e \leadsto e'$, before
translation. Importantly, our abortive handlers are more general than JAX's
\texttt{checkify}, which must complete all iterations before checking for
errors; in contrast, ours catches the exception inside the loop and aborts at
the next condition check.

These two classes do not cover every handler; for example, a clause that resumes
more than once is not supported across a loop boundary. We discuss this
limitation further in \Cref{sec:related}.

The rest of the paper proceeds as follows. \Cref{sec:calculus} presents the
calculus, \Cref{sec:effect-translation} the translation, and
\Cref{subsec:tail-resumptive-effects} its extension across loop boundaries;
all results are mechanized in Lean~4. \Cref{sec:implementation} describes the
\jaxfx{} library built on the translation, and \Cref{sec:case-studies}
evaluates it through case studies.

\section{The Calculus}
\label{sec:calculus}

This section presents \minijaxprfx{}, a calculus formalizing a core subset of
jaxprs extended with effect-handling constructs. At its heart, \minijaxprfx{} is
simply a first-order calculus with second-order primitives: functions cannot be
passed as values, but a fixed set of primitives, such as conditionals and loops,
take jaxprs as arguments. This design is not specific to JAX; rather, it
captures the essential shape of accelerator-oriented intermediate
representations. At the same time, the calculus remains faithful to actual
jaxprs, allowing our translation to apply directly to the implementation. To our
knowledge, this is the first formalization of a core jaxpr language.

\subsection{Syntax}
\label{sec:formal-syntax}

\Cref{fig:jaxpr-syntax} presents the syntax of \minijaxprfx{}. The calculus is a
statically-typed, first-order language in A-normal form
(ANF)~\cite{FlanaganANF}.
We focus on modeling a minimal set of primitives; additional primitives
can be added in a straightforward manner.
The syntax comprises three syntactic categories: types, terms (which
include expressions and jaxprs), and primitives. We assume disjoint name spaces
for variables ($x$), constants ($c$), basic primitive functions ($f$), and~effect~operations~($op$).

Types $\sigma$ are product types, written as sequences of array types for typing
tensors (i.e. multidimensional arrays). Array types $\alpha$ have the form
$d[s]$, where $d$ represents the underlying scalar data type (e.g. float, int),
and $s$ is a shape tuple specifying the number and sizes of the tensor
dimensions (e.g.\ $3 \times 4$).
For scalar values, we often omit the empty shape tuple for brevity, e.g. writing
$\booltype$ instead of $\booltype[]$.
We also often write $\unit$ for the unit type (and its unit value).

Expressions $e$ consist of a sequence of let bindings
\codejaxpr{$\deflet{b\restar}{p}{a\restar}{e}$}
that end with a return
statement $\retkw{}~ a\restar$.
Binders $b$ are typed variables. Atoms $a$ are either variables
$x$ or literals $l$ specifying tensors. Literals $l$ include constants $c$ and
tensors $[l\restar]$.
A jaxpr $j$ is a lambda expression, written \codejaxpr{$\deflambda{b\restar}{e}$}.

Primitive operations $p$ include both basic functions $f$ and higher-order
primitives $\hat{p}$ built from jaxprs. These higher-order primitives include
function calls ($\defcall{j}$), that call the jaxpr $j$ when applied within  let
bindings; conditionals ($\defcond{j_1}{j_2}$), that take a boolean argument and
execute the appropriate branch; while loops ($\defwhile{j_1}{j_2}$) with
condition jaxpr $j_1$ and  body jaxpr $j_2$; and scans
(\codejaxpr{$\defscan{j}{a_1,\dots,a_n}$}) that provide a functional loop-fold,
where $j$ iterates over input arrays while updating a \emph{carry} state
initialized by $a_1,\dots,a_n$.

To support algebraic effects, \minijaxprfx{} introduces primitives for
performing and handling effectful computations. First, the primitive
$\dokw{}~op$ performs an effectful operation $op$. Second, following
\citet{koka2017,iteratedcps2020}, we represent resumptions using a $\resumekw{}$
primitive. Finally, the syntax includes a parameterized handler
primitive: $\defhandleret{a\restar}{e}{j}{op\restar \mapsto j\restar}$.
This consists of four components:
the \emph{arguments} $a\restar$ provide the initial value for the handler
parameters; the \emph{body} expression $e$ provides the computation being handled; the \emph{return}
jaxpr $j$ converts the final handler parameters and the body's result into the
final output; and the \emph{implementation}  mapping $op\restar \mapsto j\restar$ gives
the jaxpr clauses defining the  operations.

Throughout this paper, we use an overline to denote sequences; for example,
$\as$ represents the sequence $a_1, \dots, a_n$ ($n \geq 0$), and $\as : \sigma$
indicates that each $a_i : \sigma_i$ has the corresponding type
$\sigma_i$.
\begin{figure}
  \footnotesize
  \begin{tabular}{llclllcl}
    type & $\sigma,\rho,\tau,\omega$ & $\Coloneqq$ & $\alpha\restar$ &
    expression & $e$ & $\Coloneqq$ & \deflet{b\restar}{p}{a\restar}{e} $\mid$ \retkw{} $a\restar$ \\

    array type & $\alpha$ & $\Coloneqq$ & $\dtype[\shape]$ &
    binder & $b$ & $\Coloneqq$ & $x:\alpha $ \\

    basic type & $\dtype$ & $\Coloneqq$ & $\booltype \mid \floattype \mid \inttype \mid \unittype$ \qquad &
    atom & $a$ & $\Coloneqq$ & $x \mid l$ \\

    shape &  $\shape$ & $\Coloneqq$ & $n\restar$ &
    jaxpr & $j,k$ & $\Coloneqq$ & \deflambda{b\restar}{e} \\

    & & & & literal & $l$ & $\Coloneqq$ & $c \mid [l\restar]$ \\[2pt]

    primitive & $p$ & $\Coloneqq$ & \multicolumn{5}{l}{$f \mid \hat{p} \mid \dokw{}~op \mid  \resumekw{}$}\\

    higher-order & $\hat{p}$ & $\Coloneqq$ & \multicolumn{5}{l}{$\defcall{j} \mid \defcond{j}{j} \mid \defwhile{j}{j}$} \\
     &     & $\mid$ & \multicolumn{5}{l}{$\defscan{j}{a\restar} \mid$ \defhandleret{a\restar}{e}{j}{op\restar \mapsto j\restar}} \\
  \end{tabular}

\caption{Syntax of \minijaxpr{}}
\label{fig:jaxpr-syntax}
\end{figure}
\begin{figure}[t]
\footnotesize
{\setlength{\abovedisplayskip}{1pt}%
\drules[T]{$\Env \vdash a : \sigma$}{Typing values}
{Var,Const,Array}}
\vspace{-5pt}

\begin{minipage}[t]{0.63\textwidth}
\drules[T]{$\Env \vdash e : \sigma$}{Typing expressions}
{Let,Ret}
\end{minipage}
\hfill
\begin{minipage}[t]{0.34\textwidth}
\drules[T]{$\Env \vdash j: \sigma \to \tau$}{Typing jaxprs}
{Jaxpr}
\end{minipage}
\vspace{-5pt}

\begin{drulepar}[T]{$\Env \vdash p : \sigma \to \tau$}{Typing primitives}
\drule{Func} \drule{Call} \drule{Cond} \drule{While} \drule{Scan}
\end{drulepar}
\vspace{-12pt}

\noindent
\begin{minipage}[t]{0.27\textwidth}
\vspace{0pt}
\begin{mathpar}
\ottdruleTXXDo{}
\end{mathpar}
\vspace{-5pt}
\begin{mathpar}
\ottdruleTXXResume{}
\end{mathpar}
\end{minipage}\hfill
\begin{minipage}[t]{0.71\textwidth}
\vspace{0pt}
\begin{mathpar}
\ottdruleTXXHandler{}
\end{mathpar}
\end{minipage}

\vspace{-10pt}
\caption{Typing rules for \minijaxprfx{}}
\label{fig:jaxpr-typing}
\end{figure}

\begin{figure}
  \footnotesize

  \begin{center}
    \begin{tabular}{lrcl}
      reduction & $\jaxapp{\deflambda{\xs:\sigma}{e}}{}{\as}$ & $\triangleq$ & $e[\overline{a}/\overline{x}] $ \\
      context   & $C$ & $\Coloneqq$ & $\square \mid \deflet{b\restar}{p}{a\restar}{C}$  \\
      binding   & $e_1~\bind{\bs}~ e_2$ & $\triangleq$ & $C[ e_2[\as/\bs] ]$  \quad where $e_1 = C[\retkw~\as]$ \\
    \end{tabular}
  \end{center}

  \vspace{-3pt}

{$\boxed{e \evalto e'}$ \hfill \emph{(Operational semantics)}}

\vspace{-5pt}

\renewcommand{\arraystretch}{1.05}
\setlength{\arraycolsep}{3pt}
\newcommand{\sgap}{\noalign{\vskip 3pt}}
\newcommand{\sindent}{\hspace{3em}}
\newcommand{\sspan}[1]{\multicolumn{3}{l@{}}{#1}}
\newcommand{\slhs}[1]{\sspan{#1}}
\newcommand{\srhs}[1]{\sspan{\sindent{\evalto}\;\; #1}}
\newcommand{\scont}[1]{\sspan{\sindent\phantom{\evalto\;\;} #1}}
\newcommand{\swhere}[1]{\sspan{\sindent\phantom{\evalto\;\;}\text{where } #1}}
\[
  \begin{array}{@{}llll@{}}
  \newrule{(S-Func)}
  \customlabel{S-Func}{rule:s-func}
   &
   \deflet{\bs}{f}{\as}{e} & \evalto & e[\as'/\bs] \quad \text{where } f(\as) \evalto \as' \\\sgap
  \newrule{(S-Call)}
  \customlabel{S-Call}{rule:s-call}
   &
   \deflet{\bs}{\defcall{j}}{\as}{e} & \evalto & \jaxapp{j}{}{\as}~\bind{\bs}~e \\\sgap
  \newrule{(S-Cond)}
  \customlabel{S-Cond}{rule:s-cond}
   &
   \deflet{\bs}{\defcond{j_{\true}}{j_{\false}}}{c~\as}{e} & \evalto & \deflet{\bs}{\defcall{j_c}}{\as}{e} \quad \text{where } c \in \{\true, \false\} \\\sgap
  \newrule{(S-While)}
  \customlabel{S-While}{rule:s-while}
   &
   \deflet{\xs:\sigma}{\defwhile{j_1}{j_2}}{\as}{e} & \evalto &  \\[-0.7pt]
   & \sspan{\quad
     {\arraycolsep=1pt \begin{array}[t]{@{}lll@{}}
      \multicolumn{3}{@{}l}{\deflet{z\!:\!\booltype}{\defcall{j_1}}{\as}{}} \\[-0.7pt]
      \defletnobody{\xs:\sigma}{ \mathkw{cond}\,[ & {\kww{then:}}\, \{\, \lambda \xs_1:\sigma. & \deflet{\xs_2:\sigma}{\defcall{j_2}}{\xs_1}{} }{} \\[-0.7pt]
                                                  &  &
                                                       \deflet{\xs_3:\sigma}{\defwhile{j_1}{j_2}
                                                       }{\xs_2}{\retkw~\xs_3} \} \\[-0.7pt]
                                                  &
                                                    \multicolumn{2}{l}{\kww{else:}\, \deflambda{\ys:\sigma}{\retkw~{\ys}} \,] ~ z ~\as ~
     \kw{in} ~ e }
          \end{array}}}
         \\\sgap
  \newrule{(S-Scan1)}
  \customlabel{S-Scan1}{rule:s-scan1}
   &
    \deflet{\bs}{\defscan{j}{\as}}{[l]}{e} & \evalto & \deflet{\bs}{\defcall{j}}{\as~l}{e} \\\sgap
    \newrule{(S-Scan2)}
    \customlabel{S-Scan2}{rule:s-scan2}
    & \slhs{\deflet{\xs:\sigma,y:d_2[n+1,s_2]}{\defscan{j}{\overline{a}}}{[l~\ls]}{e}} \\[-0.7pt]
    & \srhs{\deflet{\zs:\sigma,y_1:d_2[s_2]}{\defcall{j}}{\overline{a}~l}{}} \\[-0.7pt]
    & \scont{\deflet{\xs:\sigma,y_2:d_2[n,s_2]}{\defscan{j}{\ys}}{[\overline{l}]}{}} \\[-0.7pt]
    & \scont{\deflet{y:d_2[n+1,s_2]}{\cons{y_1}{y_2}}{}{e}} \\\sgap
    \newrule{(S-Hcong)}
    \customlabel{S-Hcong}{rule:s-hcong}
    & \slhs{\deflet{\xs:\sigma}{
        \defhandleret{\as_h}{e_b}{j_r} {\overline{op} \mapsto \overline{j}}
      }{\unit}{e}} \\[-0.7pt]
    & \srhs{\deflet{\xs:\sigma}{
        \defhandleret{\as_h}{e_b'}{j_r} {\overline{op} \mapsto \overline{j}}
      }{\unit}{e} \quad \text{where } e_b \evalto e_b'} \\\sgap
    \newrule{(S-Hret)}
    \customlabel{S-Hret}{rule:s-hret}
    & \slhs{\deflet{\bs}{\defhandleret{\as_h}{\retkw{}~\as}{j_r}{\overline{op} \mapsto \overline{j}}}{\unit}{e}} \\[-0.7pt]
    & \srhs{\deflet{\bs}{\defcall{j_r}}{\as_h~\as}{e}} \\\sgap
    \newrule{(S-Hdo1)}
    \customlabel{S-Hdo1}{rule:s-hdo1}
    & \slhs{\deflet{\bs}{
        \defhandleret{\as_h}{\deflet{\bs_1}{\dokw{}\, op}{\as}{e_b}}{j_r} {\overline{op} \mapsto \overline{j}}
      }{\unit}{e}} \\[-0.7pt]
    & \srhs{\deflet{\bs_1}{\dokw{} \, op}{\as}
        {
        \deflet{\bs}{
        \defhandleret{\as_h}{e_b}{j_r} {\overline{op} \mapsto \overline{j}}
        }{\unit}{e}
        }} \\[-0.7pt]
    & \swhere{op \notin \overline{op}} \\\sgap
    \newrule{(S-Hdo2)}
    \customlabel{S-Hdo2}{rule:s-hdo2}
    & \slhs{\deflet{\xs:\sigma}{
        \defhandleret{\as_h}{\deflet{\bs}{\dokw{}\, op}{\as}{e_b}}{j_r} {\overline{op} \mapsto \overline{j}}
      }{\unit}{e}} \\[-0.7pt]
    & \srhs{\deflet{\xs:\sigma}{\defcall{j[ \defcall{j_k} /\resumekw{}]}}{\as_h~\as}{e}} \\[-0.7pt]
    & \swhere{op \mapsto j \in \overline{op} \mapsto \js \land j = \deflambda{\zs:\tau,\bs'}{e'}} \\[-0.7pt]
    & \sspan{\sindent\phantom{\evalto\;\;}
         j_k = \deflambda{\zs:\tau,\bs} {
         \deflet{\ys:\sigma}{
         \defhandleret{\zs}{e_b}{j_r}{\overline{op} \mapsto \js}
         }{\unit}{\retkw~\ys}
         }} \\
  \end{array}
\]

\caption{Operational semantics for \minijaxprfx{}}
    \label{fig:minijax-rules-ext-main-text}
\end{figure}
 
\subsection{Typing}
\label{sec:type-system}

\Cref{fig:jaxpr-typing} presents the typing rules for \minijaxprfx{}.
The typing context $\Env$ is a finite sequence of variable typings
\codejaxpr{$x:\alpha$} and at most one resumption typing
\codejaxpr{$\resumekw{}$}: $\sigma \to \tau$. We assume no variable appears
twice within a context, and we work up to $\alpha$-equivalence as usual. We
use $in$ and $out$ for the input and output types of an operation, and $par$
for the type of a handler's parameter. We write $\Env \vdash \as: \sigma$ for
the component-wise derivation $\Env \vdash a_i: \sigma_i$ for all $i$
(assuming $|\as| = |\sigma|$).

There are four typing judgments; most rules are straightforward.
The judgment $\Env \vdash a : \alpha$ types atoms. In \rref{T-Const}, primitive
constants $c$ have predefined base types $d$ (e.g. $\true, \false :
\booltype$ and $\unit:\unit$). \Rref{T-Array} forms a tensor of shape $(n,
s)$ from $n$ tensors of shape $s$.
The judgment $\Env \vdash e:\sigma$ types expressions. \Rref{T-Let} checks
that the arguments $\as$ of a primitive $p : \sigma \to \tau$ have type
$\sigma$, and extends the context with $\xs:\tau$ to type the body $e$.
\Rref{T-Ret} types the returned atoms.
The judgment $\Env \vdash j : \sigma \to \tau$ types jaxprs as standard lambda
abstractions.

The judgment $\Env \vdash p : \sigma \to \tau$ types primitives. Basic
primitives $f$ (\rref{T-Basic}) are assumed to have given signatures (for
example, tensor concatenation $\;(\cons{}{})\!: \!d[s], d[ns] \to d[(n+1)s]$).
\Rref{T-Call} promotes a jaxpr to a primitive of the same type.
\Rref{T-Cond} requires both branches to have the same type $\sigma \to \tau$
and yields type $(\booltype, \sigma) \to \tau$. \Rref{T-While} follows the
standard typing of functional loops, with condition $j_1 : \sigma \to
\booltype$ and state update $j_2 : \sigma \to \sigma$. \Rref{T-Scan} takes
an initial carry of type $\sigma$ and a body $j : (\sigma, d_1[s_1]) \to
(\sigma, d_2[s_2])$, and yields type $(d_1[n,s_1]) \to (\sigma, d_2[n,s_2])$.

For effects, each operation is assumed to have a given signature $op: in \to
out$.
\Rref{T-Do} uses this signature, and requires an argument of type $in$ to
produce a result of type $out$.
\Rref{T-Handler} types handler expressions: the initial parameters $\as$ have
type $par$, the handled computation $e$ has type $\tau$, and the return clause
$j$ maps the final parameters and the result to the overall output type
$\tau'$. Each operation clause $j_i$ is typed under a context extended with
$\resumekw{}: par, out_i \to \tau'$, so that resuming takes updated
parameters and the operation result and yields the handler's output. The
handler has type $\unit \to \tau'$. \Rref{T-Resume} simply retrieves the type of
$\kw{resume}$ from the context.

\subsection{Operational Semantics}
\label{sec:op-sems}

\Cref{fig:minijax-rules-ext-main-text} presents the operational semantics of
\minijaxprfx{}, which serves as the reference behavior against which we prove
that our translation preserves semantics. The top of
\Cref{fig:minijax-rules-ext-main-text} defines auxiliary notations used in the
operational semantics. The application
$\jaxapp{\deflambda{\xs:\sigma}{e}}{}{\as}$ expands to $e[\as/\xs]$, and
$\jaxapp{\jaxapp{j}{}{\as_1}}{}{\as_2}$ abbreviates
$\jaxapp{j}{}{(\as_1,\as_2)}$. A context $C$ is either a hole $\square$ or a
sequence of let-bindings ending in a hole. Every well-formed expression can be
uniquely written as $e = C[\retkw~\as]$. The operation $e_1 ~ \bind{\bs}~ e_2$
then takes $e_1 = C[\retkw~\as]$, substitutes $\as$ for the binders $\bs$ in
$e_2$, and plugs the result into the hole of $C$. Plugging is capture-avoiding
(our mechanization uses a locally nameless representation, and our
implementation performs $\alpha$-renaming when needed). This plugging is similar to
\textit{re-normalization}, which is a well-known technique used to keep ANF
terms valid during
reduction~\cite{kennedy2007}. %

In rule \ruleref{rule:s-func}, we assume each basic primitive $f:\sigma \to \tau$ comes
with a total, deterministic evaluation relation $f(\overline{a})\evalto
\overline{a'}$ (e.g.\ $(\cons{l}{[\overline{l}]}) \to [l\overline{l}]$), and
the results are substituted for the binders $\bs$.

Rule \ruleref{rule:s-call} evaluates a call. However, reducing
$\deflet{\bs}{\defcall{j}}{\as}{e}$ directly to
$\deflet{\bs}{\jaxapp{j}{}{\as}}{\!}{e}$ would violate ANF. So instead, rule
\ruleref{rule:s-call} reduces to $\jaxapp{j}{}{\as}~\bind{\bs}~e$, which
runs the jaxpr application and binds its results in the rest of $e$.

The control-flow rules are standard. Rule \ruleref{rule:s-cond} reduces a
conditional to a call of the branch $j_c$ selected by the boolean $c$.
Rule \ruleref{rule:s-while} evaluates the condition $j_1$ and, if true, runs
one iteration of $j_2$ before calling the while primitive again; otherwise it
returns the current state $\as$. Rules \ruleref{rule:s-scan1} and
\ruleref{rule:s-scan2} evaluate a scan by induction on the input array,
running the body on the first element and recursing on the rest.

The final four rules are for handlers. Rule \ref{rule:s-hcong} evaluates the
handled computation, and rule \ref{rule:s-hret} applies the return clause to
the current parameters and the result once the computation finishes. For
operations, rule \ref{rule:s-hdo1} forwards an operation the handler does not
handle, while rule \ref{rule:s-hdo2} calls the corresponding clause $j$ with
the handler parameters and operation arguments. Within the clause,
$\kw{resume}$ is replaced by a call whose body $j_k$ captures the continuation
following the operation and reinstalls the handler. We inspect $j$ only to
determine the parameter type $\tau$ needed to construct $j_k$.
Appendix~A illustrates these rules with two example programs
that use handlers for exceptions and for state.

\subsection{Metatheory}
\label{sec:type-soundness}

We prove that \minijaxprfx enjoys the following desirable properties:

\begin{lemma}[\leanproof{Determinacy}]
  \label{lemma:determinacy}
     If\, $e \evalto e_1$ and $e \evalto e_2$, then $e_1 = e_2$.
\end{lemma}
\begin{theorem}[\leanproof]
  \label{theorem:type-soundness}
  \begin{description}
  \item[\textit{(Preservation)}]
    If\, $\Env \vdash e: \sigma$ and $e \evalto e'$ then
    $\Env \vdash e': \sigma$.
  \item[\textit{(Progress with effects)}]
    If\, $\bullet \vdash e:\sigma$, then
    $e \evalto e'$ for some $e'$, or $e$ has the form $\retkw{}~\as$, or
    $e$ has the form $\deflet{\bs}{\dokw{}~op}{\as}{e'}$.
  \end{description}
\end{theorem}

\noindent
The third case of progress arises because \minijaxprfx{} does not track
effects in types, so a well-typed program may perform an unhandled operation;
the theorem nevertheless guarantees that this is the only way it can get stuck. An effect
type system~\cite{hillerstrom2016liberating,koka2017} would rule this out, but is
orthogonal to our focus, and JAX itself does not statically track effects.

\section{Effect-Handling Translation}
\label{sec:effect-translation}

We compile effect handlers
away via a novel translation from \minijaxprfx{} to its handler-free subset
\minijaxpr{}, while leaving primitive computations to the hardware. This section
presents the core translation without control flow; control flow is addressed in
\Cref{subsec:tail-resumptive-effects}.

\begin{figure}
  \footnotesize
  \[
  \begin{array}{llll}
    \Is \enskip \Coloneqq \enskip \chandler{\bullet} \mid \chandler{\Is \cdot h} \quad &
    \chandler{h} \enskip \Coloneqq \enskip \chandler{\overline{op} \mapsto \overline{j}} \quad &
    \ps \enskip \Coloneqq \enskip \cpar{\bullet} \mid \cpar{\ps \cdot \as}  \quad &
    \ks \enskip \Coloneqq \enskip \ccont{\bullet} \mid \ccont{\ks \cdot k} \mid
                                                                                    \ccont{\ks \cdot j_r}
                                                                                    \mid \ccont{\spec{k}}
    \\
  \end{array}
  \]

  \vspace{-3pt}
{$\boxed{\newtransformeh{\Is}{\ks}{\ps}{e} : \tau_o}$ \hfill
  \emph{(Translation of expressions)}
}
\renewcommand{\arraystretch}{1.1}
\newlength{\erulew}
\settowidth{\erulew}{\newrule{(E-Hresume)}}
\newcommand{\erule}[2]{\makebox[\erulew][l]{\newrule{(#1)}}\customlabel{#1}{#2}}
\[
  \makebox[\linewidth][l]{$\begin{array}{@{}llll@{}}
  \erule{E-Ret1}{rule:e-ret1} & \newtransformeh{\bullet}{\bullet}{\bullet}{\retkw{}~\as} &
  = & \retkw ~ \as
  \\[2pt]
  \erule{E-Ret2}{rule:e-ret2}
  &
  \newtransformeh{\Is}{\ks \cdot k}{ \ps }{\retkw{}\,\as} &
  = & \newtransformeh{\Is}{\ks}{\ps}{ \jaxapp{\ccont{k}}{}{\as}}
  \\[2pt]
  \erule{E-Ret3}{rule:e-ret3}
  &
  \newtransformeh{\Is \cdot h}{\ks  \cdot j_r}{\ps \cdot \as_h  }{\retkw{}\,\as} &
  = & \newtransformeh{\Is}{\ks}{\ps}{ \jaxapp{\jaxapp{\ccont{j_r}}{}{\cpar{\as_h}}}{}{\as}}
  \\[2pt]
  \erule{E-Func}{rule:e-func}
  &
  \newtransformeh{\Is}{\ks}{\ps}{\deflet{\bs}{f}{\as}{e}} & = &
      \deflet{\bs}{f}{\as}{\newtransformeh{\Is}{\ks}{\ps }{e}}
  \\[2pt]
  \erule{E-Call1}{rule:e-call1}
  &
  \newtransformeh{\bullet}{\bullet}{\bullet}{\deflet{\bs}{\defcall{j}}{\as}{e}}
  & \\[-0.7pt]
  & & & \hspace{-135pt}
  = \; \; \deflet{\bs}{\defcall{ \newtransformjh{\bullet}{\bullet}{\bullet}{j} }}{\as}{ \newtransformeh{\bullet}{\bullet}{\bullet}{e} } \\[2pt]
  \erule{E-Call2}{rule:e-call2}
  &
  \newtransformeh{\Is}{\ks}{\ps }{\deflet{\bs}{\defcall{j}}{\as}{e}}
  &  \\[-0.7pt]
  & & & \hspace{-135pt}
        = \; \;
  \deflet{\zs: \tau_o}{\defcall{ \newtransformjh{\Is}{\ks \cdot \deflambda{\bs}{e}}{\ps  }{j} }}{\as}{\retkw~\overline{z}}
  \\[2pt]

  \erule{E-Cond1}{rule:e-cond1}
  &
  \newtransformeh{\bullet}{\bullet}{\bullet}{\deflet{\bs}
    {\defcond{j_1}{j_2}}{\as}{e}} \hspace{-30pt} & \\[-0.7pt]
  & & & \hspace{-135pt}
        = \; \;
  \deflet{\bs} {\defcond{ \newtransformjh{\bullet}{\bullet}{\bullet}{j_1} }{ \newtransformjh{\bullet}{\bullet}{\bullet}{j_2} }}{\as}{\newtransformeh{\bullet}{\bullet}{\bullet}{e}}
  \\[2pt]
  \erule{E-Cond2}{rule:e-cond2}
  &
  \newtransformeh{\Is}{\ks}{\ps }{\deflet{\bs}
  {\defcond{j_1}{j_2}}{\as}{e}} \hspace{-30pt} & \\[-0.7pt]
  & & & \hspace{-135pt}
  = \; \;
  \deflet{\overline{z}: \tau_o } {\defcond{j_1'}{j_2'}}{\as}{\retkw~\overline{z}}
  \\[-0.7pt]
  & & & \hspace{-135pt}\phantom{=\;\;\;}
    \text{where } j_1' =  \newtransformjh{\Is}{\ks \cdot \deflambda{\bs}{e}}{\ps}{j_1},
        j_2' =  \newtransformjh{\Is}{\ks \cdot \deflambda{\bs}{e}}{\ps}{j_2}
  \\[2pt]

  \erule{E-Hop}{rule:e-hop}
  &
  \newtransformeh{\Is \cdot h \cdot \Is'} {\ks \cdot j_{r} \cdot \ks'}{\ps \cdot
    \as_h \cdot \ps'}{\deflet{\xs: out}{\defdo{op}}{\as}{e}} \hspace{-120pt}
  &\\[-0.7pt]
  & & & \hspace{-135pt}
  =\;\; \newtransformeh{\Is}{\ks}{\ps}{
    \jaxapp{
      \jaxapp{j
      [ p
          / \resumekw
      ]}{}{\cpar{\as_h}}}{}{\, \as}
      } \\[-0.7pt]
  & & & \hspace{-135pt}\phantom{=\;\;\;}
      \text{where } op \mapsto j \in \chandler{h} \land op \notin \chandler{\Is'}
      \land |\chandler{\Is'}| = |\ccont{\ks'}|_{j_r} = |\cpar{\ps'}|
      \land \chandler{h}: par
  \\[-0.7pt]
  & & & \hspace{-135pt}\phantom{=\;\;\;}
      \phantom{\text{where }}
      \land p = \defcall{\deflambda{\overline{p},\xs:par,out}{\newdelay{h \cdot \Is'}{j_{r} \cdot \ks'} {\overline{p} \cdot \ps' }{e}}}
  \\[2pt]
  \erule{E-Hresume}{rule:e-hresume}
  &
    \newtransformeh{\Is}{\ks}{\ps }{
    { \newdelay{\Is'}{\ks'}{\ps'}{ e }}
  }
  &= & {\newtransformeh{\Is \cdot \Is'}{\ks \cdot \ks'}{\ps \cdot \ps'}{e}}
  \\[2pt]
  \erule{E-Handle1}{rule:e-handle1}
  &
  \newtransformeh{\bullet}{\bullet}{\bullet}{
    \deflet{\bs}{
       \defhandleret{\as_h}{e_b} {j_r}{\overline{op} \mapsto \overline{j}}
    }{\unit}{e}
  } \hspace{-150pt}
  & \\[-0.7pt]
  & & & \hspace{-135pt}
   = \;\; \newtransformeh{\overline{op} \mapsto \overline{{j}} }{j_r}{\as_h}{e_b}
      ~\bind{\bs}~
      \newtransformeh{\bullet}{\bullet}{\bullet}{e} \\[2pt]
  \erule{E-Handle2}{rule:e-handle2}
  &
  \newtransformeh{\Is}{\ks}{\ps}{
    \deflet{\bs}{
       \defhandleret{\as_h}{e_b} {j_r}{\overline{op} \mapsto \overline{j}}
    }{\unit}{e}
  } \hspace{-150pt}
  & \\[-0.7pt]
  & & & \hspace{-135pt}
   = \;\; \newtransformeh{\Is \cdot \overline{op} \mapsto \overline{{j}} }{\ks
        \cdot \deflambda{\bs}{e} \cdot j_r}{\ps \cdot \as_h}{e_b}
\end{array}$}
\]

{$\boxed{\newtransformjh{\Is}{\ks}{\ps}{j}}$ \hfill
  \emph{(Translation of jaxprs)}}

\vspace{-8pt}
\[
  \makebox[\linewidth][l]{$\begin{array}{@{}llll@{}}
  \erule{E-Jaxpr}{rule:e-jaxpr}
  &
\newtransformjh{\Is}{\ks}{\ps}{\deflambda{\xs: \rho}{e}} & = &
\deflambda{\xs: \rho}{\newtransformeh{\Is}{\ks}{\ps}{e}}
\end{array}$}
\]

\Description{\minijaxpr{} effect-handling translation. \dokw{} and \handlekw{} expressions are statically transformed away.}
\caption{Effect-handling translation}
\label{fig:effect-handling-transform-multi-nopar-nn}
\end{figure}

\subsection{The Translation}

\Cref{fig:effect-handling-transform-multi-nopar-nn} presents the translation
rules. The function $\newtransformeh{\Is}{\ks}{\ps}{e}$ translates an
expression $e$ with respect to a handler context $\Is$ mapping operations to
clauses, their parameters $\ps$, and a list of continuations $\ks$, which are
jaxprs in our setting. The list $\ks$ contains both return clauses
$\ccont{j_r}$ of handlers and normal continuations $\ccont{k}$, which we
assume are kept distinguishable. (The special form $\ccont{\spec{k}}$ is
discussed in \Cref{sec:final-cont} and can be ignored for now.)
The function
maintains the invariant that the length of $\Is$ matches both the number of
return clauses in $\ks$ and the length of $\ps$.
We assume that the output type $\tau_o$ of the elaboration is given, which
will always be provided (e.g. in \Cref{thm:nn-type-correct}).

The first three rules translate $\retkw~\as$ according to the continuation
list. If the list is empty~(\ruleref{rule:e-ret1}), the translation is
complete. Otherwise, if the innermost continuation is a normal continuation $\ccont{k}$
(\ruleref{rule:e-ret2}), we apply it to $\as$ and translate the result with
$\ccont{k}$ removed. If it is a return clause $\ccont{j_r}$
(\ruleref{rule:e-ret3}), we apply $\ccont{j_r}$ to the handler parameters
$\cpar{\as_h}$ and $\as$, and continue after popping the innermost handler
$\chandler{h}$, its parameters, and its return clause from their lists.

Primitive functions are preserved, since they are to be evaluated on the
hardware; thus, rule \ruleref{rule:e-func} keeps the function and translates the
body. For $\kw{call}$ and $\kw{cond}$, when the handler and continuation lists
are empty (\ruleref{rule:e-call1} and \ruleref{rule:e-cond1}), we simply
translate the subexpressions recursively, using the jaxpr translation given at the
bottom of the figure.
Otherwise (\ruleref{rule:e-call2} and
\ruleref{rule:e-cond2}), the subexpressions may perform operations, so we
capture the current continuation $\ccont{\deflambda{\bs}{e}}$, prepend it to the
continuation list, translate the subexpressions, apply the result to $\as$, and
return the result bound to fresh variables $\zs$.

Translating a $\kw{do}$ expression (rule \ruleref{rule:e-hop}) is slightly more
involved.
First, we look up the operation
clause $op \mapsto j$ in the handler $\chandler{h}$, together with its
corresponding parameters $\cpar{\as_h}$ and return clause $\ccont{j_r}$.
This lookup relies on the structural property that
the length of the handler list $\chandler{\Is'}$ equals both the number of return clauses
in $\ccont{\ks'}$ and the length of the parameter list $\cpar{\ps'}$.
We then translate $j$ applied to
$\cpar{\as_h}$ and $\as$ under the lists preceding $\chandler{h}$.
The complication is that $j$ may use $\kw{resume}$, which must be replaced by
its continuation. Yet we cannot translate that continuation immediately:
$\kw{resume}$ may be invoked inside $j$ under various contexts, so the context
for translating the resumed expression $e$ is unknown at this point. We
therefore substitute $\kw{resume}$ with a $\kw{call}$ primitive whose body
receives the updated parameters $\overline{p}$ and the operation results
$\xs$, and returns the special expression $\newdelay{h \cdot \Is'}{j_r \cdot
\ks'}{\overline{p} \cdot \ps' }{e}$, which captures the lists up to
$\chandler{h}$ along with the updated parameters.
Rule \ruleref{rule:e-hresume}
then resolves a $\newdelay{\Is'}{\ks'}{\ps'}{ e }$ construct by concatenating
the captured lists with the current ones and translating $e$ in the combined
context. Therefore, if $\kw{resume}$ is used several times inside $j$, $e$ is
translated under each context as intended.

Finally, rules \ruleref{rule:e-handle1} and \ruleref{rule:e-handle2}
translate $\kw{handle}$ expressions. When the handler list is empty
(\ruleref{rule:e-handle1}), we translate $e_b$ with the current handler and
bind its result to the translation of $e$ under empty lists. Otherwise
(\ruleref{rule:e-handle2}), we push the handler, its parameters, the
continuation $\ccont{\deflambda{\bs}{e}}$, and the return clause onto the
lists to translate the handled expression.

\subsection{\textbf{Elaboration Soundness and Semantics Preservation}}
\label{sec:translation-soundness}

First, it is worth noting that our translation is a \textit{selective} CPS
transformation~\cite{nielsen2001selective}:
CPS is performed only when the stack is non-empty.

\begin{lemma}[\leanproof{Pure programs}]
  \label{thm:pure}
  If \prem{e} contains no handler-related constructs, then \prem{\newtransformeh{\bullet}{\bullet}{\bullet}{e} = e}.
\end{lemma}

We then prove a group of mutually recursive structural properties of the elaboration:
\begin{lemma}[\leanproof{K-family lemmas}]
  \label{theorem:kfamily}
  \begin{description}
  \item[\textit{(K-push)}] \prem{\newtransformeh{\Is}{\ks}{\ps}{e_1~\bind{\bs}~e_2} = \newtransformeh{\Is}{\ks \cdot \deflambda{\bs}{e_2}}{\ps}{e_1}}.
  \item[\textit{(K-compose)}] \prem{\newtransformeh{\Is}{\ks \cdot \deflambda{\xs}{e_1~\bind{\bs}~e_2} \cdot \ks'}{\ps}{e} = \newtransformeh{\Is}{\ks \cdot \deflambda{\bs}{~e_2} \cdot \deflambda{\xs}{e_1} \cdot \ks'}{\ps}{e}}.
  \item[\textit{(K-delay)}] \prem{\newtransformeh{\Is}{\ks}{\ps}{\newdelay{\Is'}{\ks' \cdot \deflambda{\bs}{e_2}}{\ps'}{e_1}} = \newtransformeh{\Is}{\ks}{\ps}{\newdelay{\Is'}{\ks'}{\ps'}{e_1~\bind{\bs}~e_2}}}.
  \end{description}
\end{lemma}

We now turn to elaboration soundness. First, we define well-formedness, denoted
$\Env \vdash \Is \bowtie \ps \bowtie \ks : \tau_1 \to \tau_2$ as a judgment
indicating that the lists take an expression of type $\tau_1$ and produce one
with type $\tau_2$. Here, $\Env \vdash \chandler{h} : par \Rightarrow \tau$
abbreviates the operation-clause premises of \rref{T-Handler} for a handler that takes
parameters $par$ and has result type $\tau$.
The last premise of \ruleref{rule:wf-cont} ensures continuations always end with
a return clause.
\vspace{-2pt}
\[
  \inferrule*[right=\newrule{(WF-Empty)}\customlabel{WF-Empty}{rule:wf-empty}]{ }
  { \Env \vdash \chandler{\bullet} \bowtie \cpar{\bullet} \bowtie \ccont{\bullet} : \tau \to \tau }
  \qquad
  \inferrule*[right=\newrule{(WF-Cont)}\customlabel{WF-Cont}{rule:wf-cont}]
  { \Env \vdash \Is \bowtie \ps \bowtie \ks : \tau_2 \to \tau_3 \\
    \Env \vdash \ccont{k} : \tau_1 \to \tau_2 \\
    \Is \neq \chandler{\bullet}
  }
  { \Env \vdash \Is \bowtie \ps \bowtie \ccont{\ks \cdot k} : \tau_1 \to \tau_3 }
\]
\vspace{-10pt}
\[
  \inferrule*[right=\newrule{(WF-Handler)}\customlabel{WF-Handler}{rule:wf-handler}]
  { \Env \vdash \Is \bowtie \ps \bowtie \ks : \tau_2 \to \tau_3 \\
    \Env \vdash \chandler{h} : par \Rightarrow \tau_2 \\
    \Env \vdash \cpar{\as} : par \\
    \Env \vdash \ccont{j_r} : par, \tau_1 \to \tau_2 }
  { \Env \vdash \chandler{\Is \cdot h} \bowtie \cpar{\ps \cdot \as} \bowtie \ccont{\ks \cdot j_r} : \tau_1 \to \tau_3 }
\]

We now establish that our translation preserves typing;
a similar statement holds for jaxprs.

 \begin{restatable}[\leanproof{Elaboration Soundness}]{theorem}{preservetype}
  \label{thm:nn-type-correct}
  Given \prem{\Env \vdash \Is \bowtie \ps \bowtie \ks : \tau_1 \to \tau_{n+1}},
  if\, \prem{\Env \vdash e: \tau_1},
  then \prem{ \Env \vdash \newtransformeh{\Is}{\ks}{\ps}{e}:\tau_{n+1}}.
\end{restatable}
\vspace{-0.15\baselineskip}

Next, we prove semantics preservation.
Semantic preservation proofs usually proceed by
constructing a simulation diagram~\cite{leroy2009};
we first prove that such a diagram exists and commutes:\par
\vspace{-0.15\baselineskip}
\begin{lemma}[\leanproof{Simulation step}]
   \label{lemma:simulation}
   If \prem{e \evalto e'},
   then
   \prem{\newtransformeh{\Is}{\ks}{\ps}{e} \evalto^{*} \newtransformeh{\Is}{\ks}{\ps}{e'}}
\end{lemma}
\vspace{-0.15\baselineskip}

Using the simulation step, we then prove semantics preservation:\par
\vspace{-0.15\baselineskip}
\begin{theorem}[\leanproof{Semantics preservation}]
  \label{theorem:preserve-semantics}
  If \prem{e \evalto^{*} \retkw\,v}, then \prem{\newtransformeh{\bullet}{\bullet}{\bullet}{e} \evalto^{*} \retkw\,v}.
\end{theorem}

\section{Handling Control-Flow Primitives}
\label{subsec:tail-resumptive-effects}

The translation of the previous section does not account for control-flow
primitives such as $\kw{while}$ and $\kw{scan}$. Consider translating
{\dispsize\footnotesize $
  \newtransformeh{\Is}{\ks}{\ps}{\deflet{\bs}{\defwhile{j_1}{j_2}}{\as}{e}} $}.
Translating the loop body $j_2$ requires extending the continuation list with
the current continuation, which contains a copy of the $\kw{while}$ expression
itself. Since continuations are unrolled during translation, we
eventually return to translating the same $\kw{while}$ expression, and the
translation fails to terminate!
The root cause is that statically eliminating handlers requires statically
determining the continuation of each operation. Bounded loops such as
$\kw{scan}$ can be unrolled, but for unbounded loops no valid jaxpr
continuation can be constructed.
We are unaware of prior work addressing this challenge. Following
\Cref{sec:key-ideas}, we handle tail-resumptive handlers in-place
(\Cref{sec:resumption-inlining}) and abortive handlers by a source-level
rewrite (\Cref{sec:abortive}), and prove the tail-resumptive translation sound
(\Cref{sec:tail-soundness}).

\subsection{Handling In-Place}
\label{sec:resumption-inlining}

\begin{figure}
  \footnotesize
  \renewcommand{\arraystretch}{1.1}
  \begin{minipage}[t]{0.445\linewidth}
    \makebox[\linewidth][l]{$\begin{array}[t]{@{}l@{\hspace{7pt}}l@{}}
      \boxed{\tail{e}} &
      \tail{\deflet{\overline{x}:\sigma}{\resumekw{}}{\overline{a}}{\retkw~\overline{x}}} \\[1pt]
      &
      \tail{\deflet{\overline{x}:\sigma}{p}{\overline{a}}{\retkw~\overline{x}}} \enskip \text{if } \tail{p} \\[1pt]
      &
      \tail{\deflet{\overline{x}:\sigma}{p}{\overline{a}}{e}} \enskip \text{if } \tail{e} \,\land\, \resumekw{} \notin p \\[3pt]
      \boxed{\tail{j}} &
      \tail{\deflambda{b\restar}{e}} \enskip \text{if } \tail{e} \\[3pt]
      \boxed{\tail{p}} &
      \tail{\defcall{j}} \enskip \text{if } \tail{j} \\[1pt]
      &
      \tail{\defcond{j_1}{j_2}} \enskip \text{if } \tail{j_1} \,\land\, \tail{j_2} \\
    \end{array}$}
  \end{minipage}\hfill
  \begin{minipage}[t]{0.545\linewidth}
    \settowidth{\tailrulew}{\newrule{(R-Resume)}}
    {$\boxed{ \newTR{\Is}{\ps}{\omega}{e} }$ \hfill \emph{(Resumption inlining)}}
    \smallskip

    \makebox[\linewidth][l]{$\begin{array}[t]{@{}l@{\hspace{4pt}}l@{}}
      \trule{R-Resume}{rule:r-resume} &
      \newTR{\Is}{\ps}{\omega}{\deflet{\xs:\sigma}{\resumekw{}}{\as}{\retkw~\xs}} \\[-0.7pt]
      & \hspace{12pt} = \enskip \retkw~ \ps,\as \\[1pt]
      \trule{R-Let1}{rule:r-let1} &
      \newTR{\Is}{\ps}{\omega}{\deflet{\xs:\sigma}{p{}}{\as}{\retkw~\xs}} \\[-0.7pt]
      & \hspace{12pt} = \enskip \deflet{\ys:\ctype{\omega}}{\newTR{\Is}{\ps}{\omega}{p}{}}{\as}{\retkw~\ys}\\[1pt]
      \trule{R-Let2}{rule:r-let2} &
      \newTR{\Is}{\ps}{\tau,par,out}{\deflet{\xs:\sigma}{p}{\as}{e}} \\[-0.7pt]
      & \hspace{12pt} = \enskip
        \mathcal{E}\lBrack \, \Is \jsep \ps \jsep {\color{kcolor}\spec{\deflambda{\ys:\tau,\sigma}{ \retkw~ \ys }}} \vdash \\[-0.7pt]
      & \hspace{36pt} \deflet{\ws:\sigma}{p}{\as}{\retkw~\ws} \, \rBrack \\[-0.7pt]
      & \hspace{24pt} \bind{\zs,\xs}~{\newTR{\Is}{\zs}{\tau,par,out}{e}} \\
    \end{array}$}
  \end{minipage}
  \caption{Selected rules for resumption inlining}
  \label{fig:resumption-inlining}
  \vspace{6pt}

\end{figure}
\begin{figure}[t]
  \footnotesize
  \settowidth{\tailrulew}{\newrule{(E-While)}}
  {$\boxed{\newtransformeh{\Is}{\ks}{\ps}{e}}$
    \hfill
    \emph{(Translation of expressions)}
    }
\[
  \makebox[\linewidth][l]{$\begin{array}{@{}l@{\hspace{10pt}}l@{}}
      \rlap{\makebox[\linewidth][r]{\emph{The final continuation}}}\trule{E-Ret4}{rule:e-ret4}
      &
      \newtransformeh{\Is}{\spec{k}}{\ps}{\retkw{}\,\as} \enskip = \enskip  \jaxapp{\jaxapp{\ccont{k}}{}{\ps}}{}{\as} \\[4pt]

      \rlap{\makebox[\linewidth][r]{\emph{Handling in-place}}}\trule{E-Hop2}{rule:e-hop2}
      &
      \newtransformeh{\Is \cdot h \cdot \Is'}{\ks}{\ps \cdot \as_h \cdot \ps' }
      {\deflet{\overline{x}: out}{\defdo{op}}{\overline{a}}{e}} \\[-0.7pt]
      &  \hspace{20pt}  = \enskip  \defletnobody{\ys :\ctype{\tau,par}, \xs: out}{\defcall{{\newTR{\Is}{\ps}{\tau,par,out}{j}}}}{\cpar{\as_h}~\overline{a}} \\[-0.7pt]
      &
        \econt     \inkw~ {\newtransformeh{\Is \cdot h \cdot \Is'}{\ks}{\ys \cdot \ps'}{e}} \\[-0.7pt]
      & \econt
      \text{where } op \mapsto j \in \chandler{h} \land op \notin
         \chandler{\Is'} \land \, \tail{j} \, \land \,
         | \chandler{\Is'} | = | \cpar{\ps'} |
         \land \Is : \ctype{\tau} \land \chandler{h} : \ctype{par}
      \\[4pt]

  \rlap{\makebox[\linewidth][r]{\emph{Translating while}}}\trule{E-While}{rule:e-while1}
  &
  \newtransformeh{\Is}{\ks}{\ps}{\deflet{\xs:\sigma}{\defwhile{\deflambda{\bs_1}{e_1} }{ \deflambda{\bs_2}{e_2}
    }}{\overline{a}}{e}} \\[-0.7pt]
  & \hspace{20pt}
  = \enskip
                        \deflet{\zs:\ctype{\tau},\xs:\sigma}
                         {\defwhile{j_1}{j_2}}{\ps~\overline{a}}
                         {\newtransformeh{\Is}{\ks}{\zs}{e}}
  \\[-0.7pt]
  & \econt
                         \text{where }j_1 = \deflambda{\zs_1:\ctype{\tau},\bs_1}{\newtransformeh{\bullet}{\bullet}{\bullet}{e_1}}
  \\[-0.7pt]
  & \econt
                         \phantom{\text{where }} j_2 = \deflambda{\zs_2:\ctype{\tau},\bs_2}
                         {\newtransformeh{\Is}{\spec{\deflambda{\zs_3:\tau,\ys:\sigma}{\retkw~\zs_3, \ys}}}{\zs_2}{e_2}}
  \\
\end{array}$}
\]
\caption{Handling in-place}
  \label{fig:translate-while}

\end{figure} 
The left of \Cref{fig:resumption-inlining} defines tail-resumptive
expressions ($\tail{e}$), jaxprs ($\tail{j}$), and primitives ($\tail{p}$):
an expression is tail-resumptive if it ends in $\kw{resume}$, or if it is a
let binding whose primitive is tail-resumptive, or whose primitive contains no
$\kw{resume}$ and whose body is tail-resumptive; a jaxpr is tail-resumptive
when its body is, and $\kw{call}$ and $\kw{cond}$ are when their jaxprs are.

From a tail-resumptive expression, the judgment $\newTR{\Is}{\ps}{\omega}{e}$
extracts the pre-$\kw{resume}$ computation, which we inline instead of
materializing a continuation. It tracks the handler context $\Is$, since the
expression may perform further tail-resumptive operations, and threads the
parameters $\ps$, which such operations may update. The type
$\ctype{\omega} = \ctype{\tau,par,out}$ annotates generated let-expressions:
$\ctype{\tau}$ is the type of $\ps$, $\ctype{par}$ the parameter type of the
handler evaluated in-place, and $\ctype{out}$ the operation's output type.
We omit the straightforward rules for jaxprs and primitives. Rule
\ruleref{rule:r-resume} returns the current parameters $\ps$ alongside the
arguments to $\kw{resume}$, i.e. the updated parameters and the operation's
output. Rule \ruleref{rule:r-let1} recurses into the primitive. Rule
\ruleref{rule:r-let2} handles let bindings where body $e$ is tail-resumptive,
and primitive $p$ does not mention $\kw{resume}$. Here, the body can be
inlined recursively, and $p~\as$ is translated under $\Is$, allowing for
tail-resumptive operations of $\Is$.
Therefore, $\mathcal{R}\lBrack
\cdot \rBrack$ and $\mathcal{E}\lBrack \cdot \rBrack$ are mutually
recursive.
Notably, $p$ may install new handlers whose
operations need \textit{not} be tail-resumptive, as local handlers capture
continuations without crossing the loop boundary.

\paragraph{The final continuation}
\label{sec:final-cont}
A final subtlety of \ruleref{rule:r-let2} is that because operations of $\Is$
can update the parameters, the translation of $p~\as$ must return the final parameters.
To address this, we introduce a new continuation form $\spec{k}$, the
\textit{final continuation}, which marks the end of the computation and receives
both the current parameters and the result (as in \ruleref{rule:e-ret4}).
Accordingly, \ruleref{rule:r-let2} translates the binding under the final
continuation $\spec{\deflambda{\ys:\tau,\sigma}{\retkw~\ys}}$, so that the
updated parameters are returned, bound to $\zs$, alongside the result $\xs$, and
then translates $e$ using $\zs$ as the new parameters.

\paragraph{Handling in-place}
\label{sec:rule-handling-in-place}
Rule \ruleref{rule:e-hop2} in \Cref{fig:translate-while} handles an operation
in-place. It looks up the clause $op \mapsto j$, with $\tail{j}$, and the
parameters $\cpar{\as_h}$ in the handler $\chandler{h}$, without inspecting
the continuations, and forms a $\kw{call}$ whose body is the inlined
resumption of $j$, computed under the handlers $\Is$ preceding $\chandler{h}$,
the parameters $\ps$, and their types. The $\kw{call}$ threads the updated
parameters $\ys$ and returns the output $\xs$; the body $e$ is then translated
under the original context with the updated parameters.

Notably, this extended translation relaxes the invariant that the number of
return clauses in $\ks$ always matches the number of handlers in $\Is$ (e.g. in \ruleref{rule:e-ret4}). This is
desired: when translating $\kw{do}$, if the corresponding handler has no return
clause in $\ks$, rule \ruleref{rule:e-hop} cannot apply, leaving rule
\ruleref{rule:e-hop2} to handle only tail-resumptive operations.

\paragraph{Translation of while}
\label{subsec:while-loop-translation}

In rule \ruleref{rule:e-while1}, the loop's return type is extended with the
parameters: the translated loop takes the current parameters $\ps$ and yields
the updated ones $\zs$. The condition $j_1$ ignores the parameters and
translates $e_1$ under empty contexts\footnote{%
Since \rref{T-While} requires the condition to return only a boolean, it
  cannot return updated parameters, so the rule does not allow the condition to
  perform operations in $\Is$. This restriction can be lifted by a source-level
  rewrite that moves the condition into the loop body; see Appendix~B.1.
}; the body $j_2$ translates $e_2$ with
the input parameters, using the final continuation to return the updated
ones.

\subsection{An Example}
\label{sec:tail-correct}

\newcommand\opOne{{op_1}}
\newcommand\opTwo{{op_2}}

Resumption inlining must preserve handler scoping, which is subtle when a
handler's clause itself performs operations. Consider the following program,
where the outermost and innermost handlers handle \prem{\opOne} with clauses
$j_1$ and $j_3$, the middle one handles \prem{\opTwo}, and its \prem{\opTwo}
clause performs \prem{\opOne}:
\begin{center}
\hspace{8pt}
{\footnotesize
$\arraycolsep=0pt
 \begin{array}{lllllll}
   \kw{handle} \, [ &     \kww{args:} & \multicolumn{3}{l}{\unit, \, \kww{ret:}
                                        \, j_{r_1}, \, \kww{impl:} \,
                                        \opOne \mapsto
                                        j_1,} & & \multicolumn{1}{r@{}}{\# \text{ with } \tail{j_1}} \\[2pt]
                    &      \kww{body:} & \letkw{}~b_1 = \kw{handle}\, [ & \kww{args:} & \multicolumn{3}{l}{\unit, \, \kww{ret:} \, j_{r_2}, \,  }\\[2pt]
                    &                    &                                        & \kww{impl:} & \multicolumn{3}{l}{ \opTwo \mapsto \deflambda{\bs}{\deflet{x:\sigma}{\dokw{}~\opOne~a_1\!}{}{\deflet{y}{\resumekw}{\as}{\retkw~y}}} }, \\[2pt]
                    &                    &                                        & \kww{body:} & \letkw~b_2 = \kw{handle}\,[& \kww{args:} & \unit, \, \kww{ret:} \, j_{r_3}, \, \kww{impl:} \opOne \mapsto j_3,  \\[2pt]
                    &                    &
                    &             &                                  &
                                                                       \kww{body:}
                                                                                      &
                                                                                        \kw{let}~b_3 = \dokw{}~\opTwo~a_2 \, \kw{in} \ldots \, ] ~\unit \enskip \inkw~\ldots \,] ~\unit                   ~ \inkw{}~ \ldots \, ] ~\unit \\
 \end{array}
$
}
 \end{center}
When the innermost body performs \prem{\opTwo}, naively taking the
pre-resume part of the \prem{\opTwo} clause, i.e.\ the computation before
\kw{resume}, would yield the translation
{\dispsize\footnotesize
\[
    \kw{let}~b_3 =
    \defcall{\deflambda{\bs}{\deflet{x}{\dokw{}~\opOne~a_1\!}{}{\retkw~\as}}}~a
   \tag*{\badmark}
\]
}%
This changes the program's semantics: \prem{\opOne} is now handled by the innermost
handler ($j_3$), although it was originally performed within the scope of the
outermost handler ($j_1$). To prevent this, \citet{xie2020effect,
xie2021generalized} either restrict resumptions to be applied in the very
scope of their original handler context, or add a runtime construct that
dynamically adapts handler contexts.

Our system needs neither, as inlining naturally preserves handler scopes:
because inlining recursively processes let-bindings, it is performed under the
handler context that preceded the one being inlined, here $j_1$
(\Cref{sec:rule-handling-in-place}). Instead of the naive pre-resume part, the
inlining emits a call to the following translation,
so that \prem{\opOne} is correctly handled by $j_1$:
{\dispsize\footnotesize
\[
  \begin{array}{l}
\newTR{\opOne \mapsto j_1}{\unit}{\unit,\unit,\sigma}{\deflet{x:\sigma}{\dokw{}~\opOne~a_1\!}{}{\deflet{y}{\resumekw}{\as}{\retkw~y}}}\\
  \qquad =    \newtransformeh{\opOne \mapsto j_1}{\spec{\deflambda
        {\ys:\unit,\sigma}{ \retkw~ \ys }}}{\unit}
   {\deflet{w:\sigma}{ \dokw~\opOne~a_1\!}{}{\retkw~w}}
    \bind{z,x}~
    {\color{tcolor}\mathcal{R}\lBrack \ldots \rBrack}
\end{array}
\tag*{\okmark}
\]
}%

\subsection{Abortive Handlers}
\label{sec:rewrites}
\label{sec:abortive}

Handling in-place covers tail-resumptive handlers. Abortive handlers never
resume, so there is no resumption to inline; instead, we support them by a
source-level rewrite of the loop that leaves the translation untouched.
Concretely, we support operations $op_i$ (for $i \in 1 \ldots n$) with
signatures $in_i \rightarrow out_i$ whose abortive handlers reside outside the
loop by the following rewrite rule;
for space reasons, we give the full definitions of the
rewritten components in Appendix~B.2.
{\dispsize\footnotesize
\[\deflet{\xs: \sigma}{\defwhile{j_1}{j_2}}{\overline{a}}{e}
  \mbox{\large \quad $\leadsto$ \enskip}
  \begin{array}{l}
    \mathkw{let}\;{\zs,\mathit{i}, \overline{\overline{v}_i}: \sigma' } = {
    \defwhile{j'_1}{j'_2}}~{\overline{a}~ 0~
    \overline{\code{zeros}({in_i})} } \\
    \mathkw{in}~\deflet{\xs:\sigma}{\defgeneralcond{\overline{j_i}}}{\mathit{i}~\zs~ \overline{\overline{v}_{i}}} {e}
  \end{array}
\]
}%
\noindent
The idea is that the loop records an abortive
operation and exits, and the operation is re-performed after the loop. The
loop state carries an integer flag $i$ (initially $0$) identifying the
recorded operation, if any, together with its arguments
$\overline{\overline{v}_i}$ (initially zeros). The body $j_2'$ wraps the
original body in a handler for all $op_i$ that discards the resumption and
returns the loop state, $i$, and the operation arguments; the condition $j_1'$
additionally checks $i = 0$; and after the loop, a $\kw{cond}$ on $i$ (where
$\kww{branches}$ abbreviates nested conditions on an integer) returns the loop
results if $i = 0$ and otherwise re-performs $op_i$ with the stored arguments.
Both this rewrite and the one for effectful conditions
(Appendix~B.1) are formally proven to preserve typing and semantics; the proofs
proceed by a simulation between the source loop and the rewritten one.

\subsection{Elaboration Soundness and Semantics Preservation}
\label{sec:tail-soundness}

To show that the new translation rules
preserve elaboration soundness, we first establish the soundness of
resumption inlining:

\begin{restatable}[\leanproof{{\footnotesize$\mathcal{R}\lBrack \cdot \rBrack$} soundness}]{lemma}{tailresumptive}
  \label{lemma:tail-resumptive}
  Given \prem{\Env \vdash \ps : \overline{par_i}}
  and \prem{\Env \vdash \Is : \overline{par_i \Rightarrow \tau_i}},
  and \prem{\tail{e}}, if\, \prem{\Env, \resumekw{}: par,out \to \sigma \vdash e: \sigma},
  then \prem{ \Env \vdash \newTR{\Is}{\ps}{\overline{par_i},par,out}{e}:  \overline{par_i},par,out}.
\end{restatable}

\noindent
We then extend our well-formedness relation
$\Env \vdash \Is \bowtie \ps \bowtie \ks : \tau_1 \to \tau_2$ with a
fourth case for the final continuation $\spec{k}$:
The final continuation may only occur as the leftmost element of $\ks$:
it is the last continuation to run, as it ends the computation:
\[
  \inferrule*[right=\newrule{(WF-Final)}\customlabel{WF-Final}{rule:wf-final}]
  { \Env \vdash \Is : \overline{par_i \Rightarrow \tau_i} \\
    \Env \vdash \ps : \overline{par_i} \\
    \Env \vdash \ccont{k} : \overline{par_i}, \tau_1 \to \tau_2 }
  { \Env \vdash \Is \bowtie \ps \bowtie \spec{k} : \tau_1 \to \tau_2 }
\]
That is, $\ccont{k}$ receives the current parameters $\overline{par_i}$
alongside the result of type $\tau_1$, and
produces the final result of type $\tau_2$.
Using this extension and \Cref{lemma:tail-resumptive},
elaboration soundness (\Cref{thm:nn-type-correct}) extends to the new
translation rules.

\label{sec:tail-semantics-preservation}

Next, we turn to semantics preservation. The simulation diagram in
\Cref{sec:translation-soundness} says the following: if $e \evalto e'$, then the
translated program should take one or more steps, $\llbracket e \rrbracket
\evalto^{*} \llbracket e' \rrbracket$. For CPS transformations, however, this
diagram notoriously often doesn't commute. In particular, administrative redexes
introduced by CPS change the syntactic shape of the target program, so target
reductions may not correspond to source reductions. We avoided this issue in
\Cref{sec:translation-soundness} because there the translation generates no
administrative redexes, and when it changes the shape of a program by pushing a
continuation inward, the K-family lemmas (\Cref{theorem:kfamily}) restore
syntactic equality. Resume inlining, however, is different: rather than
following the source semantics of handlers, which resumes a captured
continuation, it executes the handler's body in place. We therefore relate the
inlined program to the CPS translation, which is already proven to match the
source semantics. Because the two differ by administrative redexes, the
simulation diagram of \Cref{sec:translation-soundness} no longer holds.
\citet{plotkin1975cps} handles this issue with a \emph{colon translation},
which contracts administrative redexes at the head of a term as it translates.
More recent work~\cite{paraskevopoulou2021cps} instead uses a logical relation to
show that, although the syntactic shape of the translated program changes, the
programs remain \emph{semantically} equivalent. Our simulation proof combines
both: a logical relation captures semantic equivalence, and, in the spirit of
Plotkin's colon translation, our simulation relation accounts for the
administrative redexes that separate the two programs.

First, we adapt this logical relation to our setting:
{\dispsize\footnotesize
\[
  \begin{array}{r@{\;}c@{\;}lr@{\;}c@{\;}l}
e_1 \refines e_2
&\enskip\triangleq\enskip&
\forall \as,\
  e_1 \evalto^{*} \retkw~ \as
  \Longrightarrow
  e_2 \evalto^{*} \retkw~ \as \qquad\qquad
&
j_1 \refines j_2
&\enskip\triangleq\enskip&
\forall \as,\
  \jaxapp{j_1}{}{\as} \refines \jaxapp{j_2}{}{\as}
\end{array}
\]}%
The expression relation says that two expressions are related
when every final value produced by the first is the same value
produced by the second. A similar relation holds for jaxprs.

\begin{lemma}[\leanproof]
  \label{lemma:lr-fundamental}
  \noindent
  \begin{description*}[mode=unboxed,font=\normalfont]
  \item[\textit{(Reflexivity)}] $e \refines e$.
  \item[\textit{(Transitivity)}] If $e_1 \refines e_2$ and $e_2 \refines e_3$,
    then $e_1 \refines e_3$.
  \end{description*}
\end{lemma}

\begin{lemma}[\leanproof]
  \label{lemma:lr-compatibility}
  \begin{description}
  \item[\textit{(Compatibility of Bind)}] If $e_1 \refines e_2$, then
    $e_1 ~\bind{\bs}~ e \;\refines\; e_2 ~\bind{\bs}~ e$.
  \item[\textit{(Compatibility of Let)}] If $e_1[\as'/\bs] \refines e_2[\as'/\bs]$ for all $\as'$,
    then $\deflet{\bs}{p}{\as}{e_1} \;\refines\; \deflet{\bs}{p}{\as}{e_2}$.
  \end{description}
\end{lemma}

\paragraph{\textbf{Simulation between translations}}

Since the tail-resumptive translation
differs syntactically from the core translation, we start by relating the two by a
simulation $\simcps{e_1} \myeqSim{j} \siminl{e_2}$ (\Cref{fig:resume-simulation})
between a program $\simcps{e_1}$ on the core side and a program $\siminl{e_2}$
on the tail-resumptive side, modulo a pending continuation $j$.
\vspace{-8pt}
\begin{figure}[H]
  \footnotesize
  {$\boxed{\simcps{e} \myeqSim{j} \siminl{e}}$ \hfill \emph{(Resume-inlining simulation)}}
  \par\vspace{-4pt}
  \renewcommand{\MathparLineskip}{\lineskiplimit=.2\baselineskip\lineskip=.2\baselineskip}%
  \begin{mathpar}
    \inferrule[\footnotesize{Sim-Resume}]{ \quad }{
      \simcps{\jaxapp{j}{}{\as}}
      \myeqSim{j}
      \siminl{\retkw~\as}
    } \rulelabel{Sim-Resume}{rule:sim-resume}
    \and
    \inferrule[\footnotesize{Sim-Admin}]{
      \simcps{\jaxapp{j_1}{}{\as}}
      \myeqSim{j}
      \siminl{e}
    }{
      \simcps{\deflet{\xs:\sigma}{\defcall{j_1}}{\as}{\retkw~\xs}}
      \myeqSim{j}
      \siminl{e}
    } \rulelabel{Sim-Admin}{rule:sim-admin}
    \and
    \inferrule[\footnotesize{Sim-Inline}]{
      \simcps{e}
      \myeqSim{j}
      \siminl{\jaxapp{j_1}{}{\as}~\bind{\xs}~e_2}
    }{
      \simcps{e}
      \myeqSim{j}
      \siminl{\deflet{\xs:\sigma}{\defcall{j_1}}{\as}{e_2}}
    } \rulelabel{Sim-Inline}{rule:sim-inline}
    \and
    \inferrule[\footnotesize{Sim-Prefix}]{
      \simcps{e_2}
      \myeqSim{j}
      \siminl{e_4}
    }{
      \simcps{e_1~\bind{\xs}~e_2}
      \myeqSim{j}
      \siminl{e_1~\bind{\xs}~e_4}
    } \rulelabel{Sim-Prefix}{rule:sim-prefix}
    \and
    \inferrule[\footnotesize{Sim-Compose}]{
      \simcps{e_2}
      \myeqSim{j}
      \siminl{e_4}
      \\
      \simcps{e_1}
      \myeqSim{\deflambda{\xs:out}{\simcps{e_2}}}
      \siminl{e_3}
    }{
      \simcps{e_1}
      \myeqSim{j}
      \siminl{e_3~\bind{\xs}~e_4}
    } \rulelabel{Sim-Compose}{rule:sim-compose}
  \end{mathpar}
  \vspace{-2pt}
  \caption{Simulation between the core translation
    (\simcps{left}) and the tail-resumptive translation (\siminl{right})}
  \label{fig:resume-simulation}
\end{figure}
The pending continuation $j$ is what the core side still runs once the
tail-resumptive side returns, so \ruleref{rule:sim-resume} relates
$\simcps{\jaxapp{j}{}{\as}}$ to $\siminl{\retkw~\as}$. The two sides may be
unfolded to different degrees: \ruleref{rule:sim-admin} lets the core side
wrap a continuation in an administrative $\kw{call}$, and
\ruleref{rule:sim-inline} lets the tail-resumptive side keep a $\kw{call}$ in
place of its inlined body. \ruleref{rule:sim-prefix} is congruence under a common
prefix, and \ruleref{rule:sim-compose} handles sequencing: when the
tail-resumptive side continues with $e_4$, the core side instead runs its
continuation $e_2$, so the prefixes are related with $\simcps{e_2}$ pending.
Like Plotkin's colon translation, this relation bridges the syntactic gap that
administrative redexes create. Plotkin's translation removes the redexes at the
head of a term; our relation keeps them and describes how the terms differ.

The relation is preserved by reduction:

\begin{lemma}[\leanproof{Step matching}]
  \label{lemma:sim-step}
  If \prem{\simcps{e_1} \myeqSim{j} \siminl{e_2}},
  and \prem{\siminl{e_2} \evalto \siminl{e_2'}},
  then there exists \prem{\simcps{e_1'}} such that
  \prem{\simcps{e_1} \evalto^{*} \simcps{e_1'}}
  and \prem{\simcps{e_1'} \myeqSim{j} \siminl{e_2'}}.
\end{lemma}

When the tail-resumptive side returns, the core side runs to the pending continuation
applied to the same atoms: if $\simcps{e_1} \myeqSim{j} \siminl{\retkw~\as}$,
then $\simcps{e_1} \evalto^{*} \jaxapp{j}{}{\as}$. At the identity continuation
$\mathit{id} = \deflambda{\xs}{\retkw~\xs}$ this is $\retkw~\as$ itself. Therefore:

\begin{lemma}[\leanproof{Adequacy of $\myeqSim{j}$}]
  \label{lemma:sim-return}
  If \prem{\simcps{e_1} \myeqSim{\mathit{id}} \siminl{e_2}},
  then \prem{\simcps{e_1} \;\refines\; \siminl{e_2}}.
\end{lemma}

\paragraph{\textbf{Semantics Preservation}}
Because resumption inlining is an instance of the simulation at the identity
continuation,
by \Cref{lemma:sim-return} we obtain:
\begin{lemma}[\leanproof{Resumption inlining preserves semantics}]
  \label{lemma:inline-resume}
  If \prem{op \mapsto j \in \chandler{h}},
  with \prem{\tail{j}},
  and \prem{\Env, \resumekw{}: par,out \to \sigma \vdash j : \sigma},
  and \prem{op \notin \chandler{\Is'}}, and
  \prem{\Env \vdash \Is' \bowtie \ps' \bowtie \ks' : \tau_1 \to \tau_2},
  then
  {\dispsize\footnotesize
  \[
    \begin{aligned}
      &\newtransformeh{\Is}{\ks}{\ps}{
        \jaxapp{\jaxapp{j[\defcall{\deflambda{\overline{p},\xs:par,out}{\newdelay{h \cdot \Is'}{j_{r} \cdot \ks'}{\overline{p} \cdot \ps'}{e_2}}}/\resumekw]}{}{\cpar{\as_h}}}{}{\,\as}} \\[2pt]
      &\qquad\refines \quad
      \newTR{\Is}{\ps}{par,out}{j}
      ~\bind{\overline{p},\xs}~
      \newtransformeh{\Is \cdot h \cdot \Is'}{\ks \cdot j_r \cdot \ks'}{\ps \cdot \overline{p} \cdot \ps'}{e_2},
    \end{aligned}
  \]}
\end{lemma}

Using \Cref{lemma:inline-resume}, we update the simulation step lemma
(\Cref{lemma:simulation}) with the logical relation:
each source step is matched up to $\refines$.
\begin{theorem}[\leanproof{Simulation step}]
  \label{theorem:sim-step-lr}
  If \prem{e \evalto e'}, then
  {\dispsize\footnotesize
    $
    \newtransformeh{\Is}{\ks}{\ps}{e'}
    \;\refines\;
    \newtransformeh{\Is}{\ks}{\ps}{e}.
    $
  }
\end{theorem}
\noindent
Since $\refines$ is transitive (\Cref{lemma:lr-fundamental}), this extends
from single steps to whole runs: a closed program and its translation return
the same values.

\begin{restatable}[\leanproof{Semantics preservation}]{theorem}{inlineagree}
  \label{theorem:inline-agree}
  If \prem{\vdash e : \tau} and \prem{e \evalto^{*} \retkw~\as}, then
  \prem{\newtransformeh{\bullet}{\bullet}{\bullet}{e} \evalto^{*} \retkw~\as}.
\end{restatable}

\section{Implementation}
\label{sec:implementation}

We implement \jaxfx{} as a JAX library, leveraging our effect-handling extension
to jaxprs.

\lstset{
  numbers=left,
  xleftmargin=2.0ex,
  basicstyle=\fontsize{6.5}{6.5}\ttfamily,
  lineskip=0pt,
}

\newlength{\LColWidth}
\setlength{\LColWidth}{.47\textwidth}
\newlength{\RColWidth}
\setlength{\RColWidth}{.47\textwidth}

\newbox\boxa
\newbox\boxc
\newbox\boxd
\newbox\boxe

\colorlet{CapBkg}{black!10}
\newcommand{\MkCap}[2]{%
  \tikz[baseline=(label.base)]{
    \node[fill=CapBkg,
          rounded corners=1pt,
          inner xsep=1.5pt,
          inner ysep=0.5pt,
          anchor=base] (label)
          {\fontsize{7.5}{8}\selectfont\textsf{#2}};
  }%
}

\newcommand{\MkPyCap}[2]{%
  \tikz[baseline=(label.base)]{
    \node[fill=CapBkg,
          rounded corners=1pt,
          inner xsep=1.5pt,
          inner ysep=0.5pt,
          anchor=base] (label)
          {\fontsize{7.5}{8}\selectfont\textsf{#2}};
  }%
}

\begin{lrbox}{\boxa}
\begin{minipage}{\LColWidth}
\begin{lstlisting}
state_ty = ShapedArray((), np.float32)
get = effect(name='get', abstract_eval=lambda: state_ty)
set = effect(name='set', abstract_eval=lambda x: ())
def double_mul(x): s = get(); set(s+s); return s*x

def get_impl(s, k): return k(s, s)
def set_impl(s, x, k): return k(x, ())
def default_impl(s, v): return (s, v)

double_mul_handled = handle(double_mul, default_impl,
  parameter=5, cases={"get": get_impl, "set": set_impl})
\end{lstlisting}
\end{minipage}
\end{lrbox}

\begin{lrbox}{\boxc}
\begin{minipage}{\RColWidth}
\begin{lstlisting}
{ lambda ; z:f32[]. let a:f32[] = delimited_handler[
   body={ lambda ; . let
     b:f32[] = effect[name=get]
     c:f32[] = add b b
     _:f32[] = effect[name=set] c
     d:f32[] = mul b z in (d,) }
   get={ lambda ; a:i32[] .
         let b:f32[] = resume a a in (b,) }
   set={ lambda ; a:i32[] b:f32[] .
         let c:f32[] = resume b b in (c,) }
   default={ lambda ; f:i32[] g:f32[]. (g,) } ] 5 in (a,) }
\end{lstlisting}
\end{minipage}
\end{lrbox}

\begin{lrbox}{\boxd}
\begin{minipage}{0.27\textwidth}
\begin{lstlisting}
{ lambda ; a:f32[] . let
    b:f32[] = closed_call[
      call_jaxpr={ lambda ; . let
          c:f32[] = add 5 5
          d:f32[] = mul 5 a
        in (d,) } ]
  in (b,) }
\end{lstlisting}
\end{minipage}
\end{lrbox}

\def\figa{\usebox\boxa}
\def\figc{\usebox\boxc}
\def\figd{\usebox\boxd}

\begin{figure}[H]
\vspace{-10pt}
\centering
\begin{minipage}[t]{0.47\textwidth}
\vspace{0pt}
\begin{subfigure}[t]{\linewidth}
\figa
\vspace{-3pt}
\caption{\jaxfx{} program}
\label{fig:effect-definition-usage}
\end{subfigure}
\end{minipage}%
\hspace{0.04\textwidth}%
\begin{minipage}[t]{0.47\textwidth}
\vspace{0pt}
\begin{subfigure}[t]{\linewidth}
\figc
\vspace{-3pt}
\caption{Traced jaxpr}
\label{fig:jaxpr-traced}
\end{subfigure}
\end{minipage}
\vspace{11pt}
\caption{An \jaxfx{} program (a) and the jaxpr it traces to (b).}
\label{fig:jaxprs}
\vspace{-1pt}
\end{figure}

\let\LColWidth\relax
\let\RColWidth\relax
\let\MkCap\relax
\let\MkPyCap\relax
 
\paragraph{\textbf{Programming with \jaxfx}}
JAX offers a NumPy-like interface~\cite{harris2020array} together with
composable transformations. \jaxfx{} extends this interface with effects and
handlers, as \Cref{fig:effect-definition-usage} illustrates. Effects are
declared with their type signatures (via \code{abstract\_eval}) and invoked like
ordinary functions, as \code{get} and \code{set} in \code{double\_mul}. A
handler is installed with \code{handle}, which takes the handled
computation, an initial \code{parameter}, a \code{default} return clause, and
the operation \code{cases}. Each case receives the handler parameter, the
operation arguments, and the continuation \code{k}, which resumes the
computation with the updated parameter and the operation's result.

\begin{wrapfigure}{r}{0.30\textwidth}
\vspace{-12pt}
\figd
\vspace{-4pt}
\caption{Compiled jaxpr}
\label{fig:jaxpr-compiled}
\end{wrapfigure}
Programs are traced into our extended jaxprs by JAX's tracing
infrastructure~\cite{frostig2019compiling}.
For example, tracing \texttt{double\_mul\_handled} yields
the jaxpr in \Cref{fig:jaxpr-traced}, which contains the effect primitives
\kw{delimited\_handler}, \kw{resume}, and \kw{effect}. Like all jaxpr
primitives, they are statically typed; each carries a \texttt{PyTree} field
describing the structure of its shape-dependent arguments.

\paragraph{\textbf{Compiling effect handlers}}
We implement the translations described in \Cref{sec:effect-translation} and
\Cref{subsec:tail-resumptive-effects}
that compile our extended jaxprs to standard JAX jaxprs.
Our implementation implements
$\alpha$-conversion and capture-avoiding substitution to systematically rename
variables and ensure freshness.
In-place handling also serves as an automated
optimization for non-iterative code. For instance, because the handlers in
\Cref{fig:jaxpr-traced} are tail-resumptive, they are inlined during the
translation to produce a more compact program shown in
\Cref{fig:jaxpr-compiled}.

\section{Case Studies}
\label{sec:case-studies}

We explore the usage of \jaxfx{} across two real-world domains in the JAX
ecosystem: numerical differential equation solving (\Cref{sec:solver}) and probabilistic
programming~(\Cref{sec:ppl}). We select these domains because they represent common JAX use
cases, both benefit from incorporating effects, and both feature mature,
established JAX libraries that serve as baselines for comparison. A
complete overview of these domains is beyond our scope; we refer interested
readers to \citet{kidger2021on} and \citet{vandemeent2018introduction} for further background.
Using these case studies, we evaluate \jaxfx{}
through the following research questions:
\begin{itemize}
\item \textbf{RQ1:} Can \jaxfx{} express effect patterns that are awkward to write in today's JAX?
\item \textbf{RQ2:} Does \jaxfx{} compose with JAX's built-in primitives, such as control flow?
\item \textbf{RQ3:} Does expressing effect patterns with \jaxfx{} come at a performance cost?
\end{itemize}

Throughout this section, all runtimes are measured on \machinespecs; compilation
(\Cref{tab:compile} and \Cref{tab:ppl-compile}) is host-side work and runs on
the machine's CPU,
\compiletimespec.

\subsection{Optimizers and Solvers}
\label{sec:solver}

Given a vector field $f$ and an initial value $y_0$, a solver
numerically approximates the trajectory $y(t)$ where $dy / dt = f(t, y)$. The
core is an \emph{integration loop}: starting from $y_0$, it repeatedly calls a
one-step method to advance the solution over time.
In
real-world solvers, the loop is a complex imperative program.
In JAX, Diffrax\footnote{\url{https://github.com/patrick-kidger/diffrax}}
implements this loop using ad-hoc mechanisms described in \Cref{subsec:motivation}.
In particular, \texttt{Diffrax}'s integration
core\footnote{\url{https://github.com/patrick-kidger/diffrax/blob/main/diffrax/_integrate.py}}
carries a single record of 20 variables serving orthogonal
purposes. The structure of this core loop is sketched
in \Cref{fig:diffrax-loop}.

We contribute a refactor of \texttt{Diffrax}'s core integration module using \jaxfx{}.

\begin{figure}[t]
\lstset{
  xleftmargin=11pt,
  numbersep=5pt,
  basicstyle=\fontsize{6.5}{6.5}\ttfamily,
  lineskip=0pt,
  commentstyle=\color{codecomment-color},
}
\begin{subfigure}[b]{0.49\textwidth}
\begin{lstlisting}
class State(eqx.Module):  # 20 fields, by hand
  y; tprev; tnext; made_jump    # evolving integration state
  solver_state; controller_state
  num_steps; num_accepted_steps # statistics + progress
  num_rejected_steps; progress_meter_state
  save_state; dense_ts          # saved + dense output
  dense_infos; dense_save_idx
  event_tprev; event_tnext      # event-detection state
  event_values; event_dense_info; event_mask
  result: RESULTS               # abort / error flag

def cond_fun(s):    # not past t1, and ok
  return (s.tprev < t1) & is_successful(s.result)

def body_fun(s):
  y, y_err, info, ss, res = solver.step(...)
  keep, tprev, tnext, ... = # thread updated state here
      stepsize_controller.adapt_step_size(...)
  ...
  # update progress meter, and thread the updated state
  progress_meter_state = progress_meter.step(...)
  return State(y=..., tprev=..., ...)  # all 20

final = lax.while_loop(cond_fun, body_fun, init)
\end{lstlisting}
  \vspace{-4pt}
\caption{\texttt{Diffrax}: manual threading, error prone.}
\label{fig:diffrax-loop}
\end{subfigure}
\hfill
\begin{subfigure}[b]{0.49\textwidth}
\begin{lstlisting}
def body_fun():   # direct style, no State
  # this step could result in an error,
  # effect automatically propagates errors and updates states
  effect.step(effect.get("tprev"), effect.get("y"), ...)
  # adapt's handlers automatically update variables
  effect.adapt(effect.get("tprev"), effect.get("y"), ...)
  # update progress meter in direct style
  effect.set("progress_meter", ...)
  ...
  _ = lax.cond(should_abort,
     lambda _: effect.fail(),
     lambda _: ())
  # return a noop, state gets automatically threaded
  return ()

def cond_fun():    # no state threading
  tprev = effect.get("tprev")
  t1 = effect.get("t1")
  return (tprev < t1)

integration_loop = lax.while_loop(
    cond_fun, body_fun, init=())
# handlers for the effects, initialize parameters
final = api.handle(..., integration_loop)

\end{lstlisting}
  \vspace{-4pt}
\caption{Diffrax with \jaxfx{}: principled, direct-style.}
\label{fig:jaxfx-loop}
\end{subfigure}

\bigskip
\vspace{3pt}
\caption{The core integration loop of \texttt{Diffrax} versus the same loop
refactored with \jaxfx{}.}
\label{fig:diffrax-comparison}
\end{figure}

\paragraph{\textbf{Expressiveness}}
With \jaxfx{}, each of these concerns becomes an independent handler and the loop
body is written in direct style. The 20-field record is replaced by one
\emph{state handler} per concern; taking a step and adapting the
step size are \emph{effects} (\texttt{try\_step},
\texttt{adapt}) interpreted by handlers; and early termination is an
\emph{exception} (\texttt{fail}) rather than a flag threaded through the state and
re-checked in the loop condition. The resulting program is written modularly in direct-style
and results in a ${\approx}22\%$ shrinkage in code size (measured by lines of code).
The same loop written in \jaxfx{} is shown in \Cref{fig:jaxfx-loop}.
With \jaxfx{}, parameterized handlers allow us to update the state in-place,
without having to worry about threading it correctly. Moreover, the loop doesn't
take in parameters itself, and the loop's conditional doesn't directly
check the results flag, rather a single \texttt{fail()} operation
immediately aborts the loop. With \jaxfx{}, the logic
can be seen directly without any obfuscation behind state threading
(\textbf{RQ1}). Note also that the effectful loop body sits inside the standard
\code{while\_loop}~and~\code{cond}~(\textbf{RQ2}).

\paragraph{\textbf{Performance}}
We compare the performance of our refactor against \texttt{Diffrax}.
From \texttt{Diffrax}'s default benchmark suite,
we select all programs that exercise the integration loop.
Each sweeps one parameter, plotted on the x-axis:
\begin{enumerate}[label=(\alph*)]
  \item a \emph{reaction-diffusion PDE} on an $N \times N$ grid, sweeping the grid
        size $N$ (increasing grid size, hence per-step work);
  \item a \emph{small neural ODE}, whose vector field is a small MLP (batch
        size 1024, i.e.\ each solver step runs the MLP on 1024 states),
        sweeping the integration length: a longer integration takes more
        solver steps, hence more work per step;
  \item a \emph{jump-step SDE} (batch size 1024, i.e.\ each solver step
        advances 1024 independent SDE paths), likewise sweeping the
        integration length;
  \item the classic \emph{Lotka-Volterra} system (a tiny two-variable ODE) as a
    \emph{work-precision}
    study~\cite{rackauckas2022ode} (increasing the error tolerance and measuring how long the solve takes).
\end{enumerate}

\definecolor{diffraxcolor}{HTML}{6395BA}   %
\definecolor{jaxfxcolor}{HTML}{F08D39}     %

\pgfplotsset{
  benchpanel/.style={
    enlarge x limits=0.08, enlarge y limits=0.12,
    tick label style={font=\scriptsize},
    label style={font=\footnotesize},
    title style={font=\footnotesize, yshift=-0.8ex},
    xtick align=outside, ytick align=outside, major tick length=2.5pt,
    xticklabel shift=-1.5pt, yticklabel shift=-1.5pt,
    axis line style={line width=0.4pt},
    every axis plot/.append style={line width=0.8pt, mark size=0.8pt},
    max space between ticks=24pt,
    xlabel shift={-3pt}, ylabel shift={-4pt}
  },
  diffraxplot/.style={diffraxcolor, mark=*},
  jaxfxplot/.style={jaxfxcolor, mark=triangle*},
  benchlegend/.style={
    legend columns=-1, legend to name=#1,
    legend style={draw=none, /tikz/every even column/.append style={column sep=1.4em}},
  },
}

\begin{figure*}[t]
\centering
\begin{minipage}{0.655\textwidth}
\centering
\begin{tikzpicture}
\begin{groupplot}[
  group style={group size=3 by 2, horizontal sep=0.95cm, vertical sep=0.5cm,
               x descriptions at=edge bottom},
  benchpanel, width=3.5cm, height=2.9cm, xmode=log, ymode=log,
  log basis x=2,
]
  \nextgroupplot[ylabel={time (ms)}, title={reaction--diffusion (PDE)},
    xtick={64,128,256,512},
    ymin=40, ymax=400,
    ytick={50,100,300}, yticklabels={50,100,300},
    benchlegend=mainleg]
  \addplot[diffraxplot] table[x=N, y=diffeqsolve, col sep=comma] {figures/eval/data/forward_against_scan.csv};
  \addlegendentry{\footnotesize\texttt{Diffrax}}
  \addplot[jaxfxplot]   table[x=N, y=diffeqsolve2, col sep=comma] {figures/eval/data/forward_against_scan.csv};
  \addlegendentry{\footnotesize\texttt{Diffrax} with \jaxfx{}}
  \nextgroupplot[title={neural ODE}, ytick={2,4,6}, yticklabels={2,4,6}]
  \addplot[diffraxplot] table[x=t1, y=diffrax, col sep=comma] {figures/eval/data/forward_small_neural_ode.csv};
  \addplot[jaxfxplot]   table[x=t1, y=diffeqsolve2, col sep=comma] {figures/eval/data/forward_small_neural_ode.csv};
  \nextgroupplot[title={jump-step (SDE)},
    ytick={200,600,1800}, yticklabels={200,600,1800}]
  \addplot[diffraxplot] table[x=t1, y=diffrax, col sep=comma] {figures/eval/data/forward_jump_step.csv};
  \addplot[jaxfxplot]   table[x=t1, y=diffeqsolve2, col sep=comma] {figures/eval/data/forward_jump_step.csv};
  \nextgroupplot[ylabel={time (ms)}, xlabel={grid $N$},
    xtick={64,128,256,512},
    ymin=300, ymax=3000,
    ytick={300,1000,3000}, yticklabels={300,1000,3000}]
  \addplot[diffraxplot] table[x=N, y=diffeqsolve, col sep=comma] {figures/eval/data/forward_grad_against_scan.csv};
  \addplot[jaxfxplot]   table[x=N, y=diffeqsolve2, col sep=comma] {figures/eval/data/forward_grad_against_scan.csv};
  \nextgroupplot[xlabel={length $t$},
    ytick={10,15,25}, yticklabels={10,15,25}]
  \addplot[diffraxplot] table[x=t1, y=diffrax, col sep=comma] {figures/eval/data/forward_grad_small_neural_ode.csv};
  \addplot[jaxfxplot]   table[x=t1, y=diffeqsolve2, col sep=comma] {figures/eval/data/forward_grad_small_neural_ode.csv};
  \nextgroupplot[xlabel={length $t$},
    ytick={500,2000,8000}, yticklabels={500,2000,8000}]
  \addplot[diffraxplot] table[x=t1, y=diffrax, col sep=comma] {figures/eval/data/forward_grad_jump_step.csv};
  \addplot[jaxfxplot]   table[x=t1, y=diffeqsolve2, col sep=comma] {figures/eval/data/forward_grad_jump_step.csv};
\end{groupplot}
\node[anchor=south] at ([yshift=-2.7mm]current bounding box.north)
  {\pgfplotslegendfromname{mainleg}};
\end{tikzpicture}
\end{minipage}%
\begin{minipage}{0.34\textwidth}
\centering
\centering
\begin{tikzpicture}
\definecolor{lvframecolor}{HTML}{7A3F6E}   %
\begin{axis}[benchpanel, width=4.6cm, height=3.1cm, xmode=log, ymode=log,
  title={Lotka--Volterra}, xlabel={error vs.\ reference (L2)}, ylabel={time (ms)},
  axis line style={line width=0.7pt, lvframecolor}]
  \addplot[diffraxplot] table[x=error, y=diffrax, col sep=comma] {figures/eval/data/lotka_volterra.csv};
  \addplot[jaxfxplot]   table[x=error, y=diffeqsolve2, col sep=comma] {figures/eval/data/lotka_volterra.csv};
\end{axis}
\end{tikzpicture}
\par\vspace{-3pt}
\pgfplotstableread[col sep=comma]{figures/eval/data/compile_times.csv}\compiletable
\pgfplotstablegetelem{0}{diffrax}\of\compiletable            \edef\DiffFwd{\pgfplotsretval}
\pgfplotstablegetelem{1}{diffrax}\of\compiletable            \edef\DiffGrad{\pgfplotsretval}
\pgfplotstablegetelem{0}{diffeqsolve2_build}\of\compiletable \edef\BuildFwd{\pgfplotsretval}
\pgfplotstablegetelem{0}{diffeqsolve2_jit}\of\compiletable   \edef\JitFwd{\pgfplotsretval}
\pgfplotstablegetelem{1}{diffeqsolve2_build}\of\compiletable \edef\BuildGrad{\pgfplotsretval}
\pgfplotstablegetelem{1}{diffeqsolve2_jit}\of\compiletable   \edef\JitGrad{\pgfplotsretval}
\pgfmathsetmacro{\PortFwd}{\BuildFwd+\JitFwd}
\pgfmathsetmacro{\PortGrad}{\BuildGrad+\JitGrad}
\newcommand{\ct}[1]{\pgfmathprintnumber[fixed, fixed zerofill, precision=2]{#1}}
\centering
{\footnotesize compile time (cold start), neural ODE}\par
\vspace{2pt}
\scriptsize
\setlength{\tabcolsep}{3.5pt}
\begin{tabular}{lcc}
\toprule
 & fwd & fwd + grad \\
\midrule
\texttt{Diffrax}  & \ct{\DiffFwd}s & \ct{\DiffGrad}s \\
with \jaxfx{} & \ct{\PortFwd}s & \ct{\PortGrad}s \\
\bottomrule
\end{tabular}
\end{minipage}
\caption{\textbf{Runtime of the \jaxfx{} refactor vs.\ \texttt{Diffrax}} (lower
is better). Left: forward solve (top) and forward solve with gradient (bottom)
on three workload benchmarks. Right: Lotka--Volterra work--precision (top) and
cold-start compilation time on \texttt{Diffrax}'s neural-ODE compile-time
benchmark (bottom).}
\label{fig:bench}
\label{tab:compile}
\end{figure*}

\Cref{fig:bench} demonstrates
the \jaxfx{} refactor is competitive on the three substantial workload benchmarks
(a)-(c) (\textbf{RQ3}).
The right panel of \Cref{fig:bench} shows the work-precision plot of the Lotka-Volterra
benchmark (d).
As the error grows ($x$-axis),
the runtime
falls ($y$-axis) since a looser tolerance does less work.
The two implementations roughly match across the entire sweep.
Lotka-Volterra is a two-variable system whose steps do minimal numerical
work, so the runtime is dominated by the integration-loop machinery itself
which is precisely where a per-step overhead of the effect-handler compilation
would show up if there were one (\textbf{RQ3}).

Finally, we compare compilation time (\Cref{fig:bench}, bottom right) from
\texttt{Diffrax}'s compile-time benchmark.
The benchmark deliberately stresses the compiler:
it is written in such a way that produces large jaxprs.
For \texttt{Diffrax}, this is the cold-start cost of the first call:
tracing its (larger) integrator program to a jaxpr and letting XLA
compile it. For \jaxfx{}, it additionally includes an ahead-of-time stage that
compiles the handlers away (tracing occurs before this step, XLA compilation after).
Despite this extra stage, the \jaxfx{} refactor compiles faster on the forward case and
matches \texttt{Diffrax} on the forward with gradient case: the effect handler compilation
pass is cheap and \texttt{Diffrax} spends longer tracing
since it has a larger program.

\subsection{Probabilistic Programming}
\label{sec:ppl}

Probabilistic models describe a joint distribution $p(x,z)$ over a space of
latent variables $z$ and observations $x$. Bayesian inference computes the
\textit{posterior} $p(z \mid x) = p(x, z) / p(x)$. The \textit{marginal
  likelihood} $p(x) = \int p(x, z) \, \mathrm{d}z$ quickly becomes intractable in
higher dimensions of latent variables, so \emph{inference algorithms} approximate the
posterior instead. \textit{Probabilistic programming languages}
(PPLs)~\cite{vandemeent2018introduction} express models with sample (drawing
from a distribution) and observe (conditioning on data) operations, and run
inference to learn model parameters from observed data.

We contribute a new PPL library in \jaxfx{} via \textit{inference
  patterns}~\cite{nguyen2023,nguyen2022modular}: an inference pattern provides \textit{inference
  operations}, whose semantics is provided by handlers, thus achieving
modularity and extensibility.

\paragraph{\textbf{Expressiveness over NumPyro and Oryx}}

We compare with two representative JAX-based PPL systems: NumPyro~\cite{phan2019numpyro} and Oryx~\cite{oryx}.
Both implement effects based on
ad-hoc mechanisms similar to those
described in \Cref{subsec:motivation}, which impose certain restrictions.
NumPyro implements effects using a global state mechanism
(``messengers'')\footnote{\url{https://num.pyro.ai/en/latest/handlers.html}}
outside of JAX, so effects break within JAX control flow.
NumPyro therefore introduces and requires users to use its own
specialized versions of \code{cond} and \code{scan},
and does not support \code{while}.
Oryx implements effects using a dedicated JAX transformation
for each effect. In
particular, its \code{harvest}
transformation\footnote{\url{https://github.com/jax-ml/oryx/tree/main/oryx/core/interpreters}}
is limited to state-like effects and currently
does not support control flow primitives.

\begin{figure}[t]
\lstset{
  numbers=left,
  xleftmargin=2.0ex,
  basicstyle=\fontsize{6.5}{6.5}\ttfamily,
  lineskip=0pt,
}

\newlength{\LColWidth}
\setlength{\LColWidth}{.500\textwidth}
\newlength{\RColWidth}
\setlength{\RColWidth}{.474\textwidth}

\newbox\boxa
\newbox\boxb
\newbox\boxc
\newbox\boxd
\newbox\boxe

\colorlet{CapBkg}{black!10}
\newcommand{\MkCap}[2]{%
  \tikz[baseline=(label.base)]{
    \node[fill=CapBkg,
          rounded corners=1pt,
          inner xsep=1.5pt,
          inner ysep=0.5pt,
          anchor=base] (label)
          {\fontsize{7.5}{8}\selectfont\textsf{#2}};
  }%
}

\begin{lrbox}{\boxa}
\begin{minipage}{\LColWidth}
\begin{lstlisting}
[@\highlightyellow{from numpyro.contrib.control\_flow import scan}@]

def gaussian_hmm(y, T: int):
  def transition(x_prev, y_curr): [@\codecomment{omitted}@]
  x0 = numpyro.sample('x_0', dist.Normal(0, 1))
  _, (x, y) = [@\highlightyellow{scan}@](transition, x0, y, length=T)
  return (x, y)[@\hfill\MkCap{numpyro}{\code{NumPyro}\protect\footnotemark{}}@]
\end{lstlisting}
\end{minipage}
\end{lrbox}

\begin{lrbox}{\boxe}
\begin{minipage}{\RColWidth}
\begin{lstlisting}
[@\highlightyellow{from jax import lax}@]

def gaussian_hmm(y, step_count: int):
  def transition(x_prev, y_curr) [@\codecomment{omitted}@]
  x0 = sample(ppl.Normal(0, 1))
  _, (x, y) = [@\highlightyellow{lax.scan}@](transition, x0, y, length=step_count)
  return (x, y)[@\hfill\MkCap{jax}{\jaxfx{}}@]
\end{lstlisting}
\end{minipage}
\end{lrbox}

\def\figa{\usebox\boxa}
\def\fige{\usebox\boxe}

\begin{minipage}{\LColWidth}
{\figa}
\end{minipage}
\hfill
\begin{minipage}{\RColWidth}
{\fige}
\end{minipage}

\caption{Hidden Markov models in NumPyro (left) and \jaxfx (right)}
\label{fig:ppl-hmm-example}

\global\let\LColWidth\relax
\global\let\RColWidth\relax
 \vspace{10pt}
\lstset{
  numbers=left,
  xleftmargin=2.0ex,
  basicstyle=\fontsize{6.5}{6.5}\ttfamily,
  lineskip=0pt,
}
\colorlet{CapBkg}{black!10}
\def\MkCap#1#2{%
  \tikz[baseline=(label.base)]{
    \node[fill=CapBkg, rounded corners=1pt, inner xsep=1.5pt,
          inner ysep=0.5pt, anchor=base] (label)
          {\fontsize{7.5}{8}\selectfont\textsf{#2}};
  }%
}

\begin{minipage}[t]{.425\textwidth}
\begin{lstlisting}
def outer(x_prev, y_row):
  x = sample("x", Normal(0.9*x_prev, 1.))
  def inner(u_prev, y):      [@\codecomment{nested}@]
    u = sample("u", Normal(0.5*u_prev+x, 0.3))
    factor(Normal(u,0.1).log_prob(y))
    return u, u
  u,_ = [@\highlightyellow{lax.scan}@](inner, 0., y_row)
  return x, u
last,_ = [@\highlightyellow{lax.scan}@](outer, 0., Y) [@\hfill\MkCap{hhmm}{\jaxfx{}}@]
\end{lstlisting}
\end{minipage}\hfill
\begin{minipage}[t]{.5350\textwidth}
\begin{lstlisting}
def step(z_prev, inp):
  r, y = inp
  z = [@\highlightyellow{lax.cond}@](r > 0,
    lambda _: sample("z", StudentT(3., z_prev, 1.)),   [@\codecomment{turbulent}@]
    lambda _: sample("z", Normal(z_prev, 0.15)), None) [@\codecomment{calm}@]
  factor(Normal(z, 0.2).log_prob(y))

  return z, z
last,_ = [@\highlightyellow{lax.scan}@](step, 0., (regimes, ys)) [@\hfill\MkCap{regime}{\jaxfx{}}@]
\end{lstlisting}
\end{minipage}

\caption{Two \jaxfx{} models that sample inside control flow: a nested
effectful \code{scan} (left) and an effectful \code{cond} (right),
neither of which NumPyro nor Oryx can express as-is.}
\label{fig:expressiveness}
\vspace{-\belowcaptionskip}
 \end{figure}
\footnotetext{Adapted from \url{https://num.pyro.ai/en/stable/primitives.html\#numpyro.contrib.control_flow.scan}.}

\Cref{fig:ppl-hmm-example} shows a \emph{Hidden Markov Model} (HMM), whose
hidden states evolve by a Markov process and are naturally expressed as a loop,
in NumPyro and in \jaxfx{} (full definitions in the artifact).
NumPyro must use its specialized \code{scan}; \jaxfx{} uses the standard JAX
\code{scan} (\textbf{RQ2}).

These restrictions can also cause a loss of expressiveness  in NumPyro and Oryx.
\Cref{fig:expressiveness} shows two practical models.
(1) A \textit{Hierarchical HMM} (left) is naturally written with nested scans.
However, NumPyro rejects nested \code{scan} with an error, and Oryx's \code{harvest}
is incompatible with JAX control flow entirely,
rejecting any latent variable inside a \code{lax.scan}.
(2) In a \textit{Regime-Switching SSM} (right), NumPyro's \code{cond} requires
all sample sites to share one distribution class and rejects the model at trace
time, while Oryx silently drops the latent variable inside \code{lax.cond} from
the joint density, giving wrong results without warning.
\jaxfx{} expresses both models directly and correctly (\textbf{RQ1},
\textbf{RQ2}).
Users can often restructure such models around these restrictions, but doing
so sacrifices compositionality and hand-encodes the error-prone interaction
between control flow and effects.

\begin{table}[t]
  \captionsetup{justification=centering}
  \begin{minipage}[t]{0.55\textwidth}
    \vspace{0pt}
\pgfplotstableread[col sep=comma]{figures/eval/data/overhead_runtime.csv}\Rtab

\newcommand{\Rr}[3]{%
  \pgfplotstablegetelem{#1}{#3}\of\Rtab\edef\tmpw{\pgfplotsretval}%
  \ifnum\tmpw=1\cellcolor{wingreen}\fi
  \pgfplotstablegetelem{#1}{#2}\of\Rtab\edef\tmpv{\pgfplotsretval}%
  \pgfmathparse{\tmpv < 100}%
  \ifdim\pgfmathresult pt>0pt
    \pgfmathprintnumber[fixed,fixed zerofill,precision=2]{\tmpv}%
  \else
    \pgfmathprintnumber[fixed,precision=0,1000 sep={}]{\tmpv}%
  \fi}

\centering
\caption{\textbf{Runtime overhead.} HW = hand-written,
  EX = \jaxfx{}, OX = Oryx, NP = NumPyro.}
\label{tab:ppl-runtime}
\footnotesize
\setlength{\tabcolsep}{3pt}
\begin{tabular}{lc||cccc}
\toprule
Model & Size & HW\,(s) & EX\,(s) & OX\,(s) & NP\,(s) \\
\midrule
Proportion      & 3M   & \Rr{0}{hand}{win_hand} & \Rr{0}{ppl2}{win_ppl2} & \Rr{0}{oryx}{win_oryx} & \Rr{0}{numpyro}{win_numpyro} \\
Stoch.\ volatility & 2.5k & \Rr{1}{hand}{win_hand} & \Rr{1}{ppl2}{win_ppl2} & \Rr{1}{oryx}{win_oryx} & \Rr{1}{numpyro}{win_numpyro} \\
Bayesian NN     & 50k  & \Rr{2}{hand}{win_hand} & \Rr{2}{ppl2}{win_ppl2} & \Rr{2}{oryx}{win_oryx} & \Rr{2}{numpyro}{win_numpyro} \\
Semi-sup.\ HMM  & 5k   & \Rr{3}{hand}{win_hand} & \Rr{3}{ppl2}{win_ppl2} & \Rr{3}{oryx}{win_oryx} & \Rr{3}{numpyro}{win_numpyro} \\
UCBadmit GLMM   & 49k  & \Rr{4}{hand}{win_hand} & \Rr{4}{ppl2}{win_ppl2} & \Rr{4}{oryx}{win_oryx} & \Rr{4}{numpyro}{win_numpyro} \\
HSGP            & 7.3k & \Rr{5}{hand}{win_hand} & \Rr{5}{ppl2}{win_ppl2} & \Rr{5}{oryx}{win_oryx} & \Rr{5}{numpyro}{win_numpyro} \\
\bottomrule
\end{tabular}
  \end{minipage}\hfill
  \begin{minipage}[t]{0.44\textwidth}
    \vspace{0pt}
\pgfplotstableread[col sep=comma]{figures/eval/data/overhead_compile.csv}\Ctab

\newcommand{\Ct}[3]{%
  \pgfplotstablegetelem{#1}{#3}\of\Ctab\edef\tmpw{\pgfplotsretval}%
  \pgfplotstablegetelem{#1}{#2}\of\Ctab\edef\tmpv{\pgfplotsretval}%
  \pgfmathsetmacro{\tmps}{\tmpv/1000}%
  \ifnum\tmpw=1\cellcolor{wingreen}\fi\pgfmathprintnumber[fixed,fixed zerofill,precision=2]{\tmps}}
\newcommand{\Cs}[2]{\pgfplotstablegetelem{#1}{#2_transform}\of\Ctab\edef\tb{\pgfplotsretval}%
  \pgfplotstablegetelem{#1}{#2_xla}\of\Ctab\edef\tx{\pgfplotsretval}%
  \pgfmathsetmacro{\tbs}{\tb/1000}\pgfmathsetmacro{\txs}{\tx/1000}%
  {\scriptsize\pgfmathprintnumber[fixed,fixed zerofill,precision=2]{\tbs}\,{+}\,\pgfmathprintnumber[fixed,fixed zerofill,precision=2]{\txs}}}

\centering
\caption{\textbf{Compile-time overhead.} Rows and
abbreviations as in \Cref{tab:ppl-runtime}.}
\label{tab:ppl-compile}
\footnotesize
\setlength{\tabcolsep}{3pt}
\begin{tabular}{ccc}
\toprule
EX\,(s) & OX\,(s) & NP\,(s) \\
\midrule
\Ct{0}{ppl2_total}{win_ppl2}\; (\Cs{0}{ppl2}) & \Ct{0}{oryx_total}{win_oryx}\; (\Cs{0}{oryx}) & \Ct{0}{numpyro_total}{win_numpyro}\; (\Cs{0}{numpyro}) \\
\Ct{1}{ppl2_total}{win_ppl2}\; (\Cs{1}{ppl2}) & \Ct{1}{oryx_total}{win_oryx}\; (\Cs{1}{oryx}) & \Ct{1}{numpyro_total}{win_numpyro}\; (\Cs{1}{numpyro}) \\
\Ct{2}{ppl2_total}{win_ppl2}\; (\Cs{2}{ppl2}) & \Ct{2}{oryx_total}{win_oryx}\; (\Cs{2}{oryx}) & \Ct{2}{numpyro_total}{win_numpyro}\; (\Cs{2}{numpyro}) \\
\Ct{3}{ppl2_total}{win_ppl2}\; (\Cs{3}{ppl2}) & \Ct{3}{oryx_total}{win_oryx}\; (\Cs{3}{oryx}) & \Ct{3}{numpyro_total}{win_numpyro}\; (\Cs{3}{numpyro}) \\
\Ct{4}{ppl2_total}{win_ppl2}\; (\Cs{4}{ppl2}) & \Ct{4}{oryx_total}{win_oryx}\; (\Cs{4}{oryx}) & \Ct{4}{numpyro_total}{win_numpyro}\; (\Cs{4}{numpyro}) \\
\Ct{5}{ppl2_total}{win_ppl2}\; (\Cs{5}{ppl2}) & \Ct{5}{oryx_total}{win_oryx}\; (\Cs{5}{oryx}) & \Ct{5}{numpyro_total}{win_numpyro}\; (\Cs{5}{numpyro}) \\
\bottomrule
\end{tabular}
  \end{minipage}
\end{table}%

\paragraph{\textbf{Performance}}

We compare \jaxfx{} against NumPyro and Oryx across both \emph{runtime} cost
and \emph{compile-time} cost using \textit{Hamiltonian Monte Carlo} (HMC)
inference. HMC is an ideal benchmark because its core loop repeatedly exercises
the target log-density gradient, which is exactly the computation generated by
each framework's side-effect abstraction,
and because it is the standard default
sampler in NumPyro.
We measure compile time for the log-density
function, and report the total runtime spent running the log-density program
over a full inference run, across $100\,000$ evaluations.
We run these benchmarks over a wide range of real-world probabilistic models
using varying large dataset sizes\footnote{\url{https://num.pyro.ai/en/0.20.0/examples/}}:
(a) proportion test logistic regression, %
(b) stochastic-volatility state space, %
(c) Bayesian neural network, %
(d) scan-marginalized semi-supervised HMM, %
(e) hierarchical GLMM based on UCBadmit dataset, %
(f) Hilbert-space GP approximation. %

\Cref{tab:ppl-runtime} reports each framework's total time
over that run, alongside a \emph{hand-written} baseline; in both
tables, the shaded cell marks the fastest entry per row. The result
shows that recovering the log-density computation via
effect handlers, \code{harvest}, or messengers adds minimal runtime
overhead, matching the baseline within a
few percent in most cases (\textbf{RQ3}).
\Cref{tab:ppl-compile} shows the compilation time.
It measures the cost of running
each library's side-effect elimination strategy to produce the log-density
computation and then XLA-compiling it. Entries
are formatted as $z\,(x + y)$ seconds, separating the framework's
effect-elimination transform time ($x$) and XLA compilation time ($y$) from
their total ($z$).
On half of the models, \jaxfx{} compiles fastest of the three frameworks, and is
within a few percent of the leader on the remaining ones (\textbf{RQ3}).

\section{Discussion and Related Work}
\label{sec:related}

\paragraph{\textbf{Compiling functional languages to accelerators}}

GPUs offer a restricted programming model and lack support for function pointers
and dynamic recursion. As a result, compiling higher-order functions typically
requires \textit{defunctionalization}~\cite{reynolds1972definitional} to produce
first-order functions, a transformation that replaces each application with a call
to an apply function that branches across all possible function forms.
However, as \citet{hovgaard2018high} observe, naive defunctionalization
introduces the issue of \textit{branch divergence} on GPUs, where extensive branching
leaves many threads idle while a subset executes instructions,
leading to suboptimal performance. To
address this, \citet{hovgaard2018high} introduced a specialized
defunctionalization transformation for the Futhark
language~\cite{henriksen2017futhark,henriksen2017design}. They restrict the use
of functions, specifically disallowing loops and conditionals from returning
functions and forbidding arrays from containing functions, a design that brings
the language remarkably close to jaxprs. Under these constraints, Futhark's
defunctionalization transformation specializes each application to the
particular form of function that may occur at runtime. As
\citet{hovgaard2018high} note, ``the result is essentially equivalent to
inlining completely the apply function in a program produced by Reynolds
defunctionalisation.''
Our translation plays the same role for effect handlers: we conjecture that it
yields the program one would obtain by CPS-transforming the handlers,
defunctionalizing the resulting higher-order functions, and fully inlining
the apply function. It does so directly, in a single pass over the first-order
program, without materializing the higher-order intermediate or inducing
branch divergence.

The array language Dex~\cite{paszke2021getting} similarly uses a simplification
pass that leaves only first-order, top-level functions that are fully applied at
their call sites, provided that the overall program is of a non-functional type.
Both Futhark and Dex normalize higher-order programs, whereas our transformation
operates on first-order jaxprs and normalizes only effect handlers.
Dex supports \texttt{Accum}, an effect for associative accumulation in parallel
computations, and later generalizes this concept to \textit{parallel effect
  handlers}~\cite{xie2024parallel}, though this feature remains yet unimplemented.
Each parallel handler must provide a \texttt{traverse} clause that defines the
semantics of operations executed in parallel. Nevertheless, this strategy
applies only to specific classes of effects, such as accumulation. Also,
their parallel exception handler must finish
parallel execution before checking for exceptions.

\paragraph{\textbf{Effect handlers.}}

Existing work on effect handler compilation
\cite{iteratedcps2020,hillerstrom2017continuation,xie2021generalized} commonly
employs a higher-order CPS transformation or targets pure lambda calculi. In
contrast, our work targets a restricted, first-order language without
first-class functions or recursion. While prior work has addressed the
mechanization of CPS compilation (e.g.~\cite{paraskevopoulou2021cps}), to our
knowledge, no prior work has mechanized a semantics-preserving CPS
transformation for effect handlers. \citet{cpseffecthandlers2022} explicitly
note that they did not mechanize their CPS semantics-preserving result as their
``mechanized translation introduces administrative reductions''.
Prior work~\cite{edward2effects-2018,nguyen2022modular,nguyen2023,
  bingham2019pyro, phan2019numpyro} has explored specialized effect handlers for
PPL, but these implementations are largely specialized to this specific domain,
while \jaxfx provides a natural interface for integrating~with~JAX.

\paragraph{\textbf{Limitations}}
We currently restrict handlers across loop boundaries to be tail-resumptive or
abortive, so a clause that e.g. resumes multiple times is rejected.
Capturing continuations in a loop requires materializing continuations as data
and turning the loop body into a state machine, following
\citet{ikebuchi2022certifying} for coroutines, though extending this approach to
multi-shot continuations remains non-trivial.
However, dispatching over states in every iteration would reintroduce the branch
divergence our translation avoids.
A further obstacle is space. Non-tail resumptive handlers require space
proportional to the number of times a loop runs. Unfortunately, this space
requirement cannot be determined at compile time and XLA requires static bounds
on memory usage for jaxprs.
A similar restriction arises in reverse-mode automatic differentiation of while
loops, which requires storing intermediate states of statically unknown memory
bounds.\footnote{\url{https://docs.jax.dev/en/latest/_autosummary/jax.lax.while_loop.html}}
We leave addressing this limitation to future work.

\paragraph{\textbf{Beyond JAX}}
Our work is motivated by the practical constraints of JAX, in the same way
that effect handler implementations for e.g. OCaml,
WebAssembly, and C/C++ are shaped by the constraints of their targets. The core
ideas, however, apply to any intermediate representation that operates under
similar hardware-imposed restrictions.
In addition to Futhark and Dex mentioned above,
PyTorch has a similar story: its
functorch component requires functional code, which PyTorch obtains through a
custom \emph{functionalization} pass\footnote{\url{https://dev-discuss.pytorch.org/t/965}}
that removes state mutation, much like JAX's \texttt{Ref} transformation. This
pass has had a number of bugs since its
introduction,\footnote{For example, \url{https://github.com/pytorch/pytorch/issues/133512}.}
and we believe effect handlers offer a
more principled treatment of side effects in this setting.
Lastly, by establishing a framework for effect handlers in JAX, we hope to open
new avenues for applying existing reasoning and verification techniques of
effect handler
programs~\cite{vilhena2021separation,biernacki2017handle,vilhena2026relational,song2022automated}
to~this~domain.

\newpage

\ifx\acksflag\paperflagtrue
\begin{acks}
  We thank the following people for their suggestions and help: Sharad Vikram,
  Dougal Maclaurin, Roy Frostig, Parker Schuh, Martin Abadi. Finally, we thank
  Dan Piponi for a prototype of effect-handling in JAX\footnote{``Handling
    Effects with
    JAX'',
    \url{https://colab.sandbox.google.com/drive/1HGs59anVC2AOsmt7C4v8yD6v8gZSJGm6}}
  that served as an inspiration for our implementation.
\end{acks}
\fi

\bibliographystyle{ACM-Reference-Format}

\appendix
\section{Example Programs}
\label{sec:source-example}

This appendix walks through two small programs, one using a handler for
exceptions and one for state, showing for each how it evaluates under the
operational semantics of \Cref{sec:calculus} and how it is translated by the
rules of \Cref{sec:effect-translation}. Throughout, $\ccont{j_r}$ denotes the
handler's return clause and $\chandler{h}$ its operation clauses.

\subsection{Exceptions}
\label{sec:source-example-exn}

Given an exception operation $\throw : \unit \to \inttype$, \safediv{} takes
two integers and divides them if the divisor is non-zero, and throws otherwise:
\[\mbox{\footnotesize\setlength{\arraycolsep}{1pt}$
\begin{array}{@{}l@{}}
  \safediv \triangleq \{\, \lambda m:\inttype, n:\inttype.\;
    \deflet{b:\booltype}{\code{eq}}{n~0}{} \\
  \hspace{2em} \mathkw{let}\; r:\inttype = \mathkw{cond}\,[\,
      \kww{then:}\, \deflambda{x:\inttype,y:\inttype}{\deflet{z:\inttype}{\dokw~\throw}{\unit}{\retkw~z}}, \\
  \hspace{8.6em} \kww{else:}\, \deflambda{x:\inttype,y:\inttype}{\deflet{z:\inttype}{\code{div}}{x~y}{\retkw~z}}
      \,]~b~m~n \\
  \hspace{2em} \mathkw{in}\; \retkw~r \,\}
\end{array}$}\]
We apply it to $42$ and $0$ under a handler whose $\throw$ clause $j$ resumes
with the fallback value $0$; the handler takes no parameter, and its return
clause $j_r$ drops the parameter:
\[\mbox{\footnotesize\setlength{\arraycolsep}{1pt}$
\begin{array}{@{}l@{}}
  \mathkw{let}\; v:\inttype = \mathkw{handle}\,[\, \kww{args:}\,\unit, \\
  \hspace{2em} \kww{body:}\, \deflet{w:\inttype}{\defcall{\safediv}}{42~0}{\retkw~w}, \\
  \hspace{2em} \kww{ret:}\, j_r = \deflambda{x:\unit,y:\inttype}{\retkw~y}, \\
  \hspace{2em} \kww{impl:}\, \throw \mapsto
      j = \deflambda{x:\unit,y:\unit}{\deflet{u:\inttype}{\resumekw}{\unit~0}{\retkw~u}} \,] ~\unit \\
  \mathkw{in}\; \retkw~v
\end{array}$}\]

\paragraph{Evaluation}
Evaluation proceeds inside the handler body (\ruleref{rule:s-hcong}). The call
is inlined with $m = 42$ and $n = 0$ (\ruleref{rule:s-call}), $\code{eq}$
evaluates to $\true$ (\ruleref{rule:s-func}), and the $\kww{then}$ branch is
selected and inlined (\ruleref{rule:s-cond}), so that, up to trivial returns,
the body becomes
$\deflet{z:\inttype}{\dokw~\throw}{\unit}{\retkw~z}$. The operation is now
performed directly under a handler that handles it, so
\ruleref{rule:s-hdo2} applies: it runs the clause $j$ with the handler
parameter and the operation argument, replacing $\kw{resume}$ by a call to the
captured continuation $j_k$, which reinstalls the handler around the rest of
the body:
\[\mbox{\footnotesize\setlength{\arraycolsep}{1pt}$
\begin{array}{@{}l@{}}
  \deflet{v:\inttype}{\defcall{j[\defcall{j_k}/\resumekw]}}{\unit~\unit}{\retkw~v} \\
  \quad\text{where } j_k = \{\, \lambda z':\unit, z:\inttype.\;
    \deflet{y:\inttype}{\defhandleret{z'}{\retkw~z}{j_r}{\throw \mapsto j}}{\unit}{} \\
  \hspace{9.5em} \retkw~y \,\}
\end{array}$}\]
Inlining $j$ (\ruleref{rule:s-call}) yields
$\deflet{u:\inttype}{\defcall{j_k}}{\unit~0}{\retkw~u}$, and inlining $j_k$
reinstalls the handler with $z = 0$. Its body has returned, so
\ruleref{rule:s-hret} applies the return clause, and \ruleref{rule:s-call}
finishes:
\[\mbox{\footnotesize\setlength{\arraycolsep}{1pt}$
\begin{array}{@{}l@{\;}l@{}}
  & \deflet{y:\inttype}{\defhandleret{\unit}{\retkw~0}{j_r}{\ldots}}{\unit}{\retkw~y} \\
  \evalto & \deflet{y:\inttype}{\defcall{j_r}}{\unit~0}{\retkw~y} \\
  \evalto^{*} & \retkw~0
\end{array}$}\]
The program's result is $0$.

\paragraph{Translation}
We translate the program under empty lists. Rule \ruleref{rule:e-handle1}
translates the handler body under the handler, with
$\chandler{h} = \chandler{\throw \mapsto j}$, continuation list
$\ccont{j_r}$, and parameter $\cpar{\unit}$, and binds the result to $v$ for
the translation of $\retkw~v$ under empty lists, which is $\retkw~v$ itself
(\ruleref{rule:e-ret1}). The interesting part is therefore
\[\mbox{\footnotesize$
  \newtransformeh{h}{j_r}{\unit}{\deflet{w:\inttype}{\defcall{\safediv}}{42~0}{\retkw~w}}.
$}\]
By \ruleref{rule:e-call2}, we capture the continuation
$\ccont{k_1} = \ccont{\deflambda{w:\inttype}{\retkw~w}}$ and translate the
jaxpr \safediv{} under $\ccont{j_r \cdot k_1}$, producing
$\deflet{z_1:\inttype}{\defcall{\newtransformjh{h}{j_r \cdot k_1}{\unit}{\safediv}}}{42~0}{\retkw~z_1}$.
Rule \ruleref{rule:e-jaxpr} recurses into the body of \safediv{};
\ruleref{rule:e-func} keeps the primitive $\code{eq}$; and
\ruleref{rule:e-cond2} captures $\ccont{k_2} = \ccont{\deflambda{r:\inttype}{\retkw~r}}$
and translates both branches under $\ccont{j_r \cdot k_1 \cdot k_2}$. The
$\kww{else}$ branch performs no operation: \ruleref{rule:e-func} keeps
$\code{div}$, and $\retkw~z$ is translated by popping $\ccont{k_2}$ and
$\ccont{k_1}$ (\ruleref{rule:e-ret2}, twice) and then $\ccont{j_r}$
(\ruleref{rule:e-ret3}), each application reducing to $\retkw~z$, so the
branch is unchanged. The $\kww{then}$ branch is where the handler is compiled
away:
\[\mbox{\footnotesize$
  \newtransformeh{h}{j_r \cdot k_1 \cdot k_2}{\unit}{\deflet{z:\inttype}{\dokw~\throw}{\unit}{\retkw~z}}.
$}\]
By \ruleref{rule:e-hop}, we look up the clause $j$ for $\throw$ in
$\chandler{h}$, replace $\kw{resume}$ in it by a call
$p = \defcall{\deflambda{p:\unit,z:\inttype}{\newdelay{h}{j_r \cdot k_1 \cdot k_2}{p}{\retkw~z}}}$
whose body delays the translation of the rest of the program, and translate
the clause applied to the handler parameter and the operation argument under
the lists preceding $\chandler{h}$, which are empty:
\[\mbox{\footnotesize\setlength{\arraycolsep}{1pt}$
\begin{array}{@{}ll@{}}
  & \newtransformeh{\bullet}{\bullet}{\bullet}{
      \jaxapp{\jaxapp{j[p/\resumekw]}{}{\cpar{\unit}}}{}{\unit}} \\
  = & \newtransformeh{\bullet}{\bullet}{\bullet}{
      \deflet{u:\inttype}{p}{\unit~0}{\retkw~u}} \\
  = & \deflet{u:\inttype}{\defcall{\deflambda{p:\unit,z:\inttype}{
      \newtransformeh{\bullet}{\bullet}{\bullet}{\newdelay{h}{j_r \cdot k_1 \cdot k_2}{p}{\retkw~z}}}}}{\unit~0}{\retkw~u} \\
  = & \deflet{u:\inttype}{\defcall{\deflambda{p:\unit,z:\inttype}{
      \newtransformeh{h}{j_r \cdot k_1 \cdot k_2}{p}{\retkw~z}}}}{\unit~0}{\retkw~u} \\
  = & \deflet{u:\inttype}{\defcall{\deflambda{p:\unit,z:\inttype}{\retkw~z}}}{\unit~0}{\retkw~u}
\end{array}$}\]
The steps are \ruleref{rule:e-call1} (the lists are empty, so the call is
kept and its body translated), \ruleref{rule:e-hresume} (the delayed
translation resumes under the captured lists), and finally the return rules
(\ruleref{rule:e-ret2} twice and \ruleref{rule:e-ret3}), which unroll
$\ccont{k_2}$, $\ccont{k_1}$, and $\ccont{j_r}$ to $\retkw~z$. The operation
has been eliminated: the call to $p$ simply yields the fallback value. The
complete translation is
\[\mbox{\footnotesize\setlength{\arraycolsep}{1pt}$
\begin{array}{@{}l@{}}
  \mathkw{let}\; z_1:\inttype = \mathkw{call}\,[\, \kww{body:}\, \{\, \lambda m:\inttype, n:\inttype.\;
    \deflet{b:\booltype}{\code{eq}}{n~0}{} \\
  \hspace{2em} \mathkw{let}\; r:\inttype = \mathkw{cond}\,[\,
      \kww{then:}\, \{\, \lambda x:\inttype,y:\inttype.\; \\
  \hspace{11em} \deflet{u:\inttype}{\defcall{\deflambda{p:\unit,z:\inttype}{\retkw~z}}}{\unit~0}{\retkw~u} \,\}, \\
  \hspace{8.6em} \kww{else:}\, \deflambda{x:\inttype,y:\inttype}{\deflet{z:\inttype}{\code{div}}{x~y}{\retkw~z}}
      \,]~b~m~n \\
  \hspace{2em} \mathkw{in}\; \retkw~r \,\} \,]~42~0 \;\; \mathkw{in}\; \retkw~z_1
\end{array}$}\]
which is a vanilla jaxpr evaluating to $0$, as the source program does.

\subsection{State}
\label{sec:source-example-state}

Consider state operations $\getop : \unit \to \inttype$ and
$\setop : \inttype \to \unit$. The handler below carries the state as its
parameter, initially $1$. Its $\getop$ clause resumes with the current state
both as the new parameter and as the operation result; its $\setop$ clause
resumes with the operation argument as the new parameter and $\unit$ as the
result; and its return clause returns the final state alongside the body's
result. The body reads the state, doubles it, stores it back, and returns
$42$:
\[\mbox{\footnotesize\setlength{\arraycolsep}{1pt}$
\begin{array}{@{}l@{}}
  \mathkw{let}\; v:\inttype,\inttype = \mathkw{handle}\,[\, \kww{args:}\, 1, \\
  \hspace{2em} \kww{body:}\, \deflet{x:\inttype}{\dokw~\getop}{\unit}{\deflet{y:\inttype}{\code{add}}{x~x}{\deflet{z:\unit}{\dokw~\setop}{y}{\retkw~42}}}, \\
  \hspace{2em} \kww{ret:}\, j_r = \deflambda{x:\inttype,y:\inttype}{\retkw~x~y}, \\
  \hspace{2em} \kww{impl:}\, \getop \mapsto
      j_g = \deflambda{x:\inttype,y:\unit}{\deflet{z:\inttype,\inttype}{\resumekw}{x~x}{\retkw~z}}, \\
  \hspace{4.7em} \setop \mapsto
      j_s = \deflambda{x:\inttype,y:\inttype}{\deflet{z:\inttype,\inttype}{\resumekw}{y~\unit}{\retkw~z}} \,] ~\unit \\
  \mathkw{in}\; \retkw~v
\end{array}$}\]

\paragraph{Evaluation}
The body immediately performs $\getop$, so \ruleref{rule:s-hdo2} runs the
clause $j_g$ with the parameter $1$ and the argument $\unit$, with
$\kw{resume}$ replaced by a call to the continuation
\[\mbox{\footnotesize\setlength{\arraycolsep}{1pt}$
\begin{array}{@{}l@{}}
  j_{k_1} = \{\, \lambda z':\inttype, x:\inttype.\;
    \deflet{y':\inttype,\inttype}{\defhandleret{z'}{e_1}{j_r}{\ldots}}{\unit}{} \\
  \hspace{6.4em} \retkw~y' \,\}, \\
  e_1 = \deflet{y:\inttype}{\code{add}}{x~x}{\deflet{z:\unit}{\dokw~\setop}{y}{\retkw~42}},
\end{array}$}\]
where $e_1$ is the rest of the body. Inlining $j_g$ and then $j_{k_1}$
(\ruleref{rule:s-call}, twice) reinstalls the handler with parameter $1$ and
$x = 1$. Inside it, $\code{add}$ evaluates to $2$ (\ruleref{rule:s-func},
under \ruleref{rule:s-hcong}), and the body performs $\setop~2$. Again
\ruleref{rule:s-hdo2} runs $j_s$ with parameter $1$ and argument $2$; its
$\kw{resume}~2~\unit$ reinstalls the handler with parameter $2$ around the
remaining body $\retkw~42$. That body has returned, so \ruleref{rule:s-hret}
applies $j_r$ to $2$ and $42$, and the program evaluates to $2, 42$: the final
state and the body's result.

\paragraph{Translation}
As before, \ruleref{rule:e-handle1} translates the body under
$\chandler{h} = \chandler{\getop \mapsto j_g, \setop \mapsto j_s}$, the
continuation list $\ccont{j_r}$, and the parameter $\cpar{1}$, and binds the
result to $v$. The body starts with $\getop$, so \ruleref{rule:e-hop} applies:
$\kw{resume}$ in $j_g$ is replaced by
$p_1 = \defcall{\deflambda{p:\inttype,x:\inttype}{\newdelay{h}{j_r}{p}{e_1}}}$,
and $j_g[p_1/\resumekw]$ applied to $\cpar{1}$ and $\unit$ is translated under
empty lists. By \ruleref{rule:e-call1} the call is kept and its body
translated, and by \ruleref{rule:e-hresume} that body becomes
$\newtransformeh{h}{j_r}{p}{e_1}$: the rest of the body, now translated with
the \emph{updated} parameter $p$ in place of $1$. Rule \ruleref{rule:e-func}
keeps $\code{add}$, and the $\setop$ operation is handled by
\ruleref{rule:e-hop} in the same way, with
$p_2 = \defcall{\deflambda{p':\inttype,z:\unit}{\newdelay{h}{j_r}{p'}{\retkw~42}}}$
and the clause $j_s[p_2/\resumekw]$ applied to $p$ and $y$. Its delayed body
$\newtransformeh{h}{j_r}{p'}{\retkw~42}$ is resolved by
\ruleref{rule:e-ret3}: the return clause is applied to the current parameter
$p'$ and the result, giving $\retkw~p'~42$. Assembling the pieces, the
complete translation is
\[\mbox{\footnotesize\setlength{\arraycolsep}{1pt}$
\begin{array}{@{}l@{}}
  \mathkw{let}\; z:\inttype,\inttype = \mathkw{call}\,[\, \kww{body:}\,
    \{\, \lambda p:\inttype, x:\inttype.\;
      \deflet{y:\inttype}{\code{add}}{x~x}{} \\
  \hspace{2em}
      \deflet{z':\inttype,\inttype}{\defcall{\deflambda{p':\inttype, z'':\unit}{\retkw~p'~42}}}{y~\unit}{} \\
  \hspace{2em} \retkw~z' \,\} \,]~1~1 \;\; \mathkw{in}\; \retkw~z
\end{array}$}\]
Both handlers and both operations are gone. The state is threaded explicitly:
the outer call receives the initial state $1$ twice, as the parameter and as
the result of $\getop$; the inner call receives the doubled value $2$ as the
new parameter, which the inlined return clause pairs with $42$. The jaxpr
evaluates to $2, 42$, as the source program does.
 \section{Source-Level Rewrites of While Loops}
\label{sec:appendix-rewrites}
\subsection{Effectful Conditions}
\label{sec:cond-rewrite}

The translation of $\kw{while}$ in \Cref{fig:translate-while} forbids
operations of outer handlers in the loop condition. This restriction can be
lifted by moving the condition into the loop body, with an additional boolean
carry $r$, through the source-level rewrite
{\footnotesize
\[
  \deflet{\xs:\sigma}{\defwhile{j_1}{j_2}}{\overline{a}}{e}
  \mbox{\large \enskip $\rwcond$ \enskip}
  \deflet{r:\booltype,\, \xs:\sigma}{\defwhile{j'_1}{j'_2}}{\true~\overline{a}}{e}
\]}%
with the rewritten condition $j_1'$ and body $j_2'$ defined as follows:
{\footnotesize
  \[\arraycolsep=1pt
    \begin{array}{rclll}
      j'_1 &=& \multicolumn{3}{l}{\deflambda{r:\booltype,\, \ys:\sigma}{\retkw~r}}\\[2pt]
      j'_2 &=& \text{\{\,\lambdasym{}} r:\booltype,\, \ys:\sigma\text{.} & \deflet{b:\booltype}{\defcall{j_1}}{\ys}{} \hspace{-120pt}\\
           & & & \defletnobody{b':\booltype,\, \zs:\sigma}{\mathkw{cond}\,[}{\!\!}
           & \kww{then:}\,
             \deflambda{\zs:\sigma}{\deflet{\ws:\sigma}{\defcall{j_2}}{\zs}{\retkw~\true~\ws}}
      \\
           & & & &\kww{else:}\, \deflambda{\zs:\sigma}{\retkw~\false~\ys} \;] ~ b~\ys \\
           & & &  \kw{in} ~\retkw~b'~\zs\;\}\\[8pt]
    \end{array}\]}%

\noindent
The new condition $j_1'$ returns the boolean carry $r$. The new body $j_2'$
evaluates the original condition $j_1$ on the state $\ys$ and branches on the
result: if true, it runs the original body $j_2$ and returns $\true$ together
with the new state; otherwise it returns $\false$ with the current state,
which makes $j_1'$ terminate the loop at the next iteration.
\subsection{Abortive Handlers}
\label{sec:abortive-rewrite}

Abortive handlers never resume, so an operation performed inside a loop and
handled abortively outside it needs no continuation: the loop can simply stop
and the operation be re-performed after it. \Cref{sec:abortive} supports this
by a source-level rewrite, applied before translation, for operations $op_i$
(for $i \in 1 \ldots n$) with signatures $in_i \rightarrow out_i$ whose
abortive handlers reside outside the loop:
{\footnotesize
\[\deflet{\xs: \sigma}{\defwhile{j_1}{j_2}}{\overline{a}}{e}
  \mbox{\large \quad $\rwabort$ \enskip}
  \begin{array}{l}
    \mathkw{let}\;{\zs,\mathit{i}, \overline{\overline{v}_i}: \sigma' } = {
    \defwhile{j'_1}{j'_2}}~{\overline{a}~ 0~
    \overline{\code{zeros}({in_i})} } \\
    \mathkw{in}~\deflet{\xs:\sigma}{\defgeneralcond{\overline{j_i}}}{\mathit{i}~\zs~ \overline{\overline{v}_{i}}} {e}
  \end{array}
\]}%
The loop state is extended with an integer flag $i$, initially $0$, that
records which operation aborted the loop, if any, together with slots for its
arguments, initially zeros. After the loop, a $\kw{cond}$ on the flag either
returns the loop results or re-performs the recorded operation with the stored
arguments, where $\mathkw{branches}$ abbreviates nested conditions on an
integer. The rewritten components are defined as follows:
{\footnotesize\arraycolsep=1pt
  \[\begin{array}{lcl}
     \sigma' &=& \sigma, \inttype, \overline{\overline{in}_{i}}\\
     j'_1 &=& \text{\{\,\lambdasym{}} \zs, \mathit{i}, \overline{\overline{v}_{i}}: \sigma'.\\
          & & \hspace{10pt} \deflet{b_1:\booltype}{\defcall{j_1}}{\zs}{}\\
          & & \hspace{10pt} \deflet{b_2:\booltype}{\code{eq}}{\mathit{i}~0}{}\\
          & & \hspace{10pt} \deflet{b:\booltype}{\code{and}}{b_1~b_2}{\retkw~b ~\}}\\[2pt]
     j'_2 &=& \{ \, \lambda \zs, \mathit{i}, \overline{\overline{v}_i}: \sigma'.~ \\
          & & \hspace{10pt}
              \mathkw{let}~ \ws : \sigma' = \mathkw{handle}\,[\,\kww{args:}\,\unit,\,\kww{body:}\,\jaxapp{j_2}{}{\zs},\,\\
          & & \hspace{15pt}
              \kww{ret:}\, \deflambda{\ys:\sigma}{\retkw~\ys~ 0 ~ \overline{\code{zeros}({in_i})}}, \\[1pt]
          & & \hspace{15pt} \kww{impl:}\, \overline{op_i} \mapsto
              \overline{ \{\, \lambda {\overline{v}: in_i}. } \\
          & & \hspace{20pt}
              \overline{\retkw~\zs~i~\code{zeros}(\overline{in_{\ldots
              i-1}})~\overline{v} ~ \code{zeros}(\overline{in_{i+1 \ldots}}) }
              \;\}  ~] ~\unit \\
         & &  \hspace{10pt}
               \mathkw{in}~ \retkw~\ws ~\}\\[2pt]
     j_0 &=& \deflambda{\ys, \mathit{i}, \overline{\overline{v}_{i}}: \sigma'}{\retkw~\ys}\\[2pt]
     j_i &=&  \{\, \lambda {\ys, \mathit{i}, \overline{\overline{v}_{i}}: \sigma'}.~
           \deflet{\zs:out_i}{\dokw{}\, op_i}{\overline{v}_{i}}{\retkw~\zs} ~ \}
\end{array}\]}%

\noindent
The extended state type $\sigma'$ adds to the original loop state $\sigma$ an
integer flag $i$ and the argument types $\overline{in_i}$ of every operation.
The condition $j_1'$ evaluates the original condition $j_1$ on the original
state $\zs$ and conjoins it with the check $i = 0$. The body $j_2'$ runs the
original body $j_2$ under a handler with a unit parameter. Its return clause
passes the updated state through with $i = 0$ and zeroed argument slots. Its
clause for $op_i$ discards the resumption and returns the current state $\zs$,
the identifier $i$, and the operation arguments $\overline{v}$ placed in the
$i$-th slot, with all other slots zeroed. Finally, the branch jaxprs
$\overline{j_i}$ are selected on the flag after the loop: $j_0$ returns the
loop results, while $j_i$ (for $i \geq 1$) re-performs $op_i$ with the stored
arguments $\overline{v}_i$.
\subsection{Elaboration Soundness and Semantics Preservation}
\label{sec:rewrite-soundness}

We write $\rwcond$ for the effectful conditions rewrite of
\Cref{sec:cond-rewrite} and $\rwabort$ for the abortive handlers rewrite of
\Cref{sec:abortive-rewrite}. We prove that both rewrites preserve typing.
\begin{theorem}[\leanproof{Effectful Conditions Elaboration Soundness}]
  \label{theorem:rewrite-sound-cond}
  If \prem{\Env \vdash e_1 : \tau} and \prem{e_1 \rwcond e_2}, then \prem{\Env \vdash e_2 : \tau}.
\end{theorem}

\begin{theorem}[\leanproof{Abortive Handlers Elaboration Soundness}]
  \label{theorem:rewrite-sound-abortive}
  If \prem{\Env \vdash e_1 : \tau} and \prem{e_1 \rwabort e_2}, then \prem{\Env \vdash e_2 : \tau}.
\end{theorem}

\noindent
The rewrites also preserve semantics: whenever the original loop returns, the
rewritten loop returns the same atoms.
For the rewrite in \Cref{sec:cond-rewrite}, we first start developing
a syntactic relation (on the source program) that captures how the
rewrite changes the syntax of the program ($\condsim$).
We then show that this relation is closed
under step:
\begin{lemma}[\leanproof{Effectful Conditions Step Simulation}]
  \label{lemma:esim-step}
  If \prem{\Env \vdash e_1 : \tau}, \prem{e_1 \condsim e_2}, and
  \prem{e_1 \evalto e'_1}, then \prem{e_2 \evalto^{*} e'_2} for some $e'_2$
  with \prem{e'_1 \condsim e'_2}.
\end{lemma}

\noindent Using \Cref{lemma:esim-step}, we then show semantics preservation:
\begin{theorem}[\leanproof{Effectful Conditions Semantics Preservation}]
  \label{theorem:rewrite-sem-cond}
  If \prem{\Env \vdash e_1 : \tau}, \prem{e_1 \rwcond e_2}, and \prem{e_1 \evalto^{*} \retkw~\as}, then
  \prem{e_2 \evalto^{*} \retkw~\as}.
\end{theorem}

For the semantics preservation of \Cref{sec:abortive-rewrite},
we also develop a syntactic relation on the source program that shows
how the rewrite changes the program shape ($\abortsim$). We then prove
step simulation:
\begin{lemma}[\leanproof{Abortive Handlers Step Simulation}]
  \label{lemma:asim-step}
  If \prem{e_1 \abortsim e_2} and \prem{e_1 \evalto e'_1}, then
  \prem{e_2 \evalto^{*} e'_2} for some $e'_2$ with \prem{e'_1 \abortsim e'_2}.
\end{lemma}

\noindent
Using \Cref{lemma:asim-step}, we can prove semantics preservation.
Since the abortive rewrite is a peephole rewrite, thus such a contextual
program could end in an unhandled operation. This operation could be
an abortive do or not. The theorem therefore says that if they
the step results in a do then the continuaiton of the do should
also be in the relation, if it is not abortive. If it is abortive, it
doesn't matter because the continuation is aborted. 
\begin{theorem}[\leanproof{Abortive Handlers Semantics Preservation}]
  \label{theorem:rewrite-sem-abortive}
  If \prem{\Env \vdash e_1 : \tau} and \prem{e_1 \rwabort e_2}, then
  \begin{description}
  \item[\textit{(Returns)}] if \prem{e_1 \evalto^{*} \retkw~\as}, then
    \prem{e_2 \evalto^{*} \retkw~\as};
  \item[\textit{(Operations)}] if
    \prem{e_1 \evalto^{*} \deflet{\bs}{\dokw{}~op}{\as}{e}}, then
    \prem{e_2 \evalto^{*} \deflet{\bs}{\dokw{}~op}{\as}{e'}};
    if $op \notin A$, then \prem{e \abortsim e'}.
  \end{description}
\end{theorem}

The abortive rewrite applies to a loop nested anywhere in a program, for
instance inside the body of another loop or of a handler. We therefore state
its semantics preservation for the abortive rewrite congruence
$e_1 \rwcong e_2$ of \Cref{fig:abort-rewrite-cong}, which applies $\rwabort$
at any number of positions in $e_1$. The judgment
is parameterized by the set $A$ of abortive operations, whose handlers never
resume, and \rref{Rw-Loop} only rewrites loops whose $op_i$ are in $A$.
\Rref{Rw-Loop} states the rewrite of a loop $\deflet{\xs:\sigma}{\defwhile{j_1}{j_2}}{\as}{\retkw~\xs}$ without its
continuation: the rewritten loop ends in the dispatch $e_r$ on the flag, which
is then composed with the original continuation $e$. The remaining rules are
congruences. \Rref{Rw-Call} rewrites a called jaxpr; the rules for the
jaxprs of $\kw{cond}$, $\kw{while}$ and $\kw{scan}$ are similar.
\rref{Rw-Handle} rewrites the body of an abortive handler $h$, i.e.\ one
whose clauses never resume.
\rref{Rw-Handle} also has other variant which can rewrite
the return and operation clauses of such handlers.

\begin{figure}[H]
  \footnotesize
  {$\boxed{e \rwcong e'}$ \hfill \emph{(Abortive rewrite congruence)}}
  \par\vspace{-4pt}
  \renewcommand{\MathparLineskip}{\lineskiplimit=.8\baselineskip\lineskip=.8\baselineskip}
  \begin{mathpar}
    \inferrule[\footnotesize{Rw-Loop}]{
      \begin{array}{l}
        \deflet{\xs:\sigma}{\defwhile{j_1}{j_2}}{\as}{\retkw~\xs} \cr
        \quad\rwabort
        \deflet{\ys:\sigma'}{\defwhile{j'_1}{j'_2}}{\as'}{e_r}
      \end{array}
      \\ \overline{op_i} \in A
    }{
      \begin{array}{l}
        \deflet{\xs:\sigma}{\defwhile{j_1}{j_2}}{\as}{e} \cr
        \quad\rwcong
        \deflet{\ys:\sigma'}{\defwhile{j'_1}{j'_2}}{\as'}{(e_r~\bind{\xs}~e)}
      \end{array}
    } \rulelabel{Rw-Loop}{rule:rw-loop}
    \and
    \inferrule[\footnotesize{Rw-Let}]{
      e \rwcong e'
    }{
      \deflet{\xs:\sigma}{p}{\as}{e}
      \rwcong
      \deflet{\xs:\sigma}{p}{\as}{e'}
    } \rulelabel{Rw-Let}{rule:rw-let}
    \and
    \inferrule[\footnotesize{Rw-Call}]{
      j \rwcong j'
    }{
      \deflet{\xs:\sigma}{\defcall{j}}{\as}{e}
      \rwcong
      \deflet{\xs:\sigma}{\defcall{j'}}{\as}{e}
    } \rulelabel{Rw-Call}{rule:rw-call}
    \and
    \inferrule[\footnotesize{Rw-Handle}]{
      e_1 \rwcong e'_1
      \\ h \text{ is abortive}
    }{
      \begin{array}{l}
        \deflet{\xs:\sigma}{\defhandleret{\as}{e_1}{j_r}{h}}{}{e} \cr
        \quad\rwcong
        \deflet{\xs:\sigma}{\defhandleret{\as}{e'_1}{j_r}{h}}{}{e}
      \end{array}
    } \rulelabel{Rw-Handle}{rule:rw-handle}
  \end{mathpar}
  \caption{Abortive rewrite congruence (selected rules).}
  \label{fig:abort-rewrite-cong}
\end{figure}

\noindent
Using the lemmas \Cref{lemma:asim-step,theorem:rewrite-sem-abortive},
we then prove semantics preservation for the contexual equivalence:
\begin{theorem}[\leanproof{Abortive Rewrite Congruence Semantics Preservation}]
  \label{theorem:rewrite-sem-cong}
  If \prem{\vdash e_1 : \tau}, \prem{e_1 \rwcong e_2}, and
  \prem{e_1 \evalto^{*} \retkw~\as}, then \prem{e_2 \evalto^{*} \retkw~\as}.
\end{theorem}
 \section{Mechanization}
\label{sec:appendix-mechanization}
\Cref{tab:lean} lists the Lean counterpart of each result in the paper.
\begin{table}[h]
\centering
\caption{Paper results and their Lean counterparts. Files are \texttt{JaxEffects/\textit{File}.lean}.}
\label{tab:lean}
\scriptsize
\setlength{\tabcolsep}{3pt}
\begin{tabular}{llllr}
\toprule
Result & Paper Name & Lean Name & File & Line \\
\midrule
\Cref{lemma:determinacy} & Determinacy & \texttt{determinacy} & \texttt{MiniJaxprH} & 2976 \\
\Cref{theorem:type-soundness} & Preservation & \texttt{preservation} & \texttt{MiniJaxprH} & 8419 \\
\Cref{theorem:type-soundness} & Progress with effects & \texttt{progress} & \texttt{MiniJaxprH} & 8130 \\
\midrule
Definition & $\mathcal{E}\lBrack\cdot\rBrack$ & \texttt{translateExpr} & \texttt{Translation} & 213 \\
Definition & $\mathcal{J}\lBrack\cdot\rBrack$ & \texttt{translateJaxpr} & \texttt{Translation} & 456 \\
\Cref{thm:pure} & Pure programs & \texttt{pure\_programs} & \texttt{Kfamily} & 13 \\
\Cref{theorem:kfamily} & K-push & \texttt{K\_push} & \texttt{Kfamily} & 3418 \\
\Cref{theorem:kfamily} & K-compose & \texttt{K\_compose} & \texttt{Kfamily} & 3465 \\
\Cref{theorem:kfamily} & K-delay & \texttt{K\_delay} & \texttt{Kfamily} & 3518 \\
\Cref{thm:nn-type-correct} & Elaboration soundness & \texttt{elaboration\_soundness} & \texttt{ElaborationSoundness} & 1070 \\
\Cref{lemma:simulation} & Simulation step & \texttt{simulation\_step} & \texttt{SemanticsPreservation} & 3200 \\
\Cref{theorem:preserve-semantics} & Semantics preservation & \texttt{semantics\_preservation} & \texttt{SemanticsPreservation} & 4711 \\
\midrule
Definition & $\mathcal{R}\lBrack\cdot\rBrack$ & \texttt{RExpr} & \texttt{Translation} & 463 \\
\Cref{lemma:tail-resumptive} & $\mathcal{R}\lBrack\cdot\rBrack$ soundness & \texttt{resumption\_inlining\_typed} & \texttt{ElaborationSoundness} & 1365 \\
Definition & Logical relation & \texttt{Refines} & \texttt{Simulation} & 121 \\
\Cref{lemma:lr-fundamental} & Reflexivity & \texttt{Refines\_refl} & \texttt{Simulation} & 124 \\
\Cref{lemma:lr-fundamental} & Transitivity & \texttt{Refines\_trans} & \texttt{Simulation} & 127 \\
\Cref{lemma:lr-compatibility} & Compatibility (Bind) & \texttt{Refines\_bind} & \texttt{Simulation} & 305 \\
\Cref{lemma:lr-compatibility} & Compatibility (Let) & \texttt{Refines\_let} & \texttt{Simulation} & 2455 \\
Definition & Simulation relation & \texttt{Rsim}$_1$ & \texttt{Simulation} & 409 \\
Definition & Simulation relation (top level) & \texttt{Rsim}$_0$ & \texttt{Simulation} & 456 \\
\Cref{lemma:sim-step} & Step matching & \texttt{sim\_step} & \texttt{Simulation} & 1595 \\
\Cref{lemma:sim-return} & Adequacy of $\myeqSim{j}$ & \texttt{simulation\_preserves\_semantics} & \texttt{Simulation} & 2372 \\
\Cref{lemma:inline-resume} & Resumption inlining preserves semantics & \texttt{resumption\_inlining\_preserves\_semantics} & \texttt{SemanticsPreservation} & 1951 \\
\Cref{theorem:sim-step-lr} & Simulation step & \texttt{simulation\_step} & \texttt{SemanticsPreservation} & 3200 \\
\Cref{theorem:inline-agree} & Semantics preservation & \texttt{semantics\_preservation} & \texttt{SemanticsPreservation} & 4711 \\
\midrule
\Cref{theorem:rewrite-sound-cond} & Effectful conditions: typing & \texttt{translateWhile\_preserves\_typing} & \texttt{EffectfulConditional} & 436 \\
Definition & $\condsim$ & \texttt{WSim} & \texttt{EffectfulConditional} & 784 \\
\Cref{lemma:esim-step} & Effectful conditions: step & \texttt{WSim\_step} & \texttt{EffectfulConditional} & 1242 \\
\Cref{theorem:rewrite-sem-cond} & Effectful conditions: semantics & \texttt{translateWhile\_semantics\_preservation} & \texttt{EffectfulConditional} & 1756 \\
\Cref{theorem:rewrite-sound-abortive} & Abortive handlers: typing & \texttt{translateWhile\_preserves\_typing} & \texttt{Abortive} & 1619 \\
Definition & $\abortsim$ & \texttt{ASimCong} & \texttt{Abortive} & 5998 \\
\Cref{lemma:asim-step} & Abortive handlers: step & \texttt{ASimCong\_step} & \texttt{Abortive} & 6904 \\
\Cref{theorem:rewrite-sem-abortive} & Abortive handlers: semantics & \texttt{translateWhile\_preserves\_terminals} & \texttt{Abortive} & 5858 \\
Definition & $\rwcong$ & \texttt{RewriteExprAnywhere} & \texttt{Abortive} & 508 \\
\Cref{theorem:rewrite-sem-cong} & Rewrite congruence: semantics & \texttt{RewriteExprAnywhere\_star} & \texttt{Abortive} & 7522 \\
\bottomrule
\end{tabular}
\end{table}
\end{document}